\documentstyle[11pt,amssymb]{article}

\makeatletter
\@addtoreset{equation}{section}
\makeatother

\def\dalemb#1#2{{\vbox{\hrule height .#2pt
        \hbox{\vrule width.#2pt height#1pt \kern#1pt
                \vrule width.#2pt}
        \hrule height.#2pt}}}

\def\cA{{\cal A}}
\def\B{{\cal B}}

\def\hi{\hat{\imath}}
\def\hj{\hat{\jmath}}

\def\0{{\sst{(0)}}}
\def\1{{\sst{(1)}}}
\def\2{{\sst{(2)}}}
\def\3{{\sst{(3)}}}
\def\4{{\sst{(4)}}}
\def\5{{\sst{(5)}}}
\def\6{{\sst{(6)}}}
\def\7{{\sst{(7)}}}
\def\8{{\sst{(8)}}}
\def\n{{\sst{(n)}}}

\def\Z{\rlap{\sf Z}\mkern3mu{\sf Z}}

\def\G{{\cal G}}

\def\CC{{\cal C}}
\def\S{{\cal S}}

\def\cD{{\cal D}}

\def\cO{{\cal O}}
\def\cL{{\cal L}}

\let\a=\alpha  \let\g=\gamma \let\d=\delta \let\e=\epsilon
 \let\h=\eta \let\q=\theta  \let\k=\kappa
\let\l=\lambda \let\m=\mu \let\n=\nu  \let\p=\pi \let\r=\rho
\let\s=\sigma \let\t=\tau   \let\c=\chi 
\let\w=\omega  \let\D=\Delta  
      \let\vart=\vartheta
    \let\T=\Theta  \let\G=\Gamma  \let\vare=\varepsilon 
   \let\Si=\Sigma
   
\def\nn{\nonumber} \def\bd{\begin{document}} \def\ed{\end{document}}
\def\ds{\documentstyle} \let\fr=\frac \let\bl=\bigl \let\br=\bigr
\def\HW{Ho\v{r}ava-Witten }
\def\WZW{Wess-Zumino-Witten }
\let\Br=\Bigr \let\Bl=\Bigl 
\let\na=\nabla
\let\pa=\partial \let\ov=\overline 
\newcommand{\be}{\begin{equation}} 
\newcommand{\ee}{\end{equation}} 
\def\ba{\begin{array}}
\def\ea{\end{array}}
\def\ft#1#2{{\textstyle{{\scriptstyle #1}\over {\scriptstyle #2}}}}
\def\fft#1#2{{#1 \over #2}}
\def\del{\partial}
\def\sst#1{{\scriptscriptstyle #1}}
\def\oneone{\rlap 1\mkern4mu{\rm l}}
\def\ie{{\it i.e.\ }}
\def\etc{{\it etc.\ }}
\def\via{{\it via}}
\def\semi{{\ltimes}}
\def\cV{{\cal V}}
\def\str{{\rm str}}
\def\jm{{\rm j}}
\def\im{{\rm i}}
\def\mapright#1{\smash{\mathop{-\!\!\!-\!\!\!-\!\!\!-\!\!\!-\!\!\!
             \longrightarrow}\limits^{#1}}}
\def\maprightt#1#2{\smash{\mathop{-\!\!\!-\!\!\!-\!\!\!-\!\!\!-\!\!\!
             \longrightarrow}\limits^{#1}_{#2}}}
\def\bb#1{{\mbox{\boldmath $#1$}}}

\def\cramp{\medmuskip = 2mu plus 1mu minus 2mu}
\def\cramper{\medmuskip = 2mu plus 1mu minus 2mu}
\def\crampest{\medmuskip = 1mu plus 1mu minus 1mu}
\def\uncramp{\medmuskip = 4mu plus 2mu minus 4mu}

\newcommand{\ho}[1]{$\, ^{#1}$}
\newcommand{\hoch}[1]{$\, ^{#1}$}
\newcommand{\bea}{\begin{eqnarray}} 
\newcommand{\eea}{\end{eqnarray}} 
\newcommand{\ra}{\rightarrow}
\newcommand{\lra}{\longrightarrow}
\newcommand{\Lra}{\Leftrightarrow}
\newcommand{\ap}{\alpha^\prime}
\newcommand{\bp}{\tilde \beta^\prime}
\newcommand{\tr}{{\rm tr} }
\newcommand{\Tr}{{\rm Tr} } 
\newcommand{\NP}{Nucl. Phys. }
\newcommand{\tamphys}{\it Center for Theoretical Physics\\
Texas A\&M University, College Station, Texas 77843}
\newcommand{\ens}{\it Laboratoire de Physique Th\'eorique de l'\'Ecole
Normale Sup\'erieure\hoch{2,3}\\
24 Rue Lhomond - 75231 Paris CEDEX 05}
\newcommand{\upenn}{\it David Rittenhouse Laboratory\\
Department of Physics and Astronomy\\
University of Pennsylvania, Philadelphia, Pennsylvania 19104}
\newcommand{\ufc}{\it Departamento de Fisica\\
Universidade Federal do Cear\'a, Fortaleza, Brazil}

\newcommand{\auth}{Eduardo Lima\hoch{1,2}, Burt Ovrut\hoch{1},
Jaemo Park\hoch{1} and Ren\'{e} Reinbacher\hoch{1\dagger}}

\def\hk{\hat{k}}
\def\hM{\hat{M}}
\def\hN{\hat{N}}
\def\hO{\hat{O}}
\def\hP{\hat{P}}
\def\hA{\hat{A}}
\def\hB{\hat{B}}
\def\hC{\hat{C}}
\def\hE{\hat{E}}
\def\ha{\hat{a}}
\def\hb{\hat{b}}
\def\hc{\hat{c}}
\def\hbZ{\hat{\Bbb{Z}}}
\def\hbA{\hat{\Bbb{A}}}
\def\hbC{\hat{\Bbb{C}}}
\def\hbB{\hat{\Bbb{B}}}
\def\hbD{\hat{\Bbb{D}}}
\def\hbM{\hat{\Bbb{M}}}
\def\hbN{\hat{\Bbb{N}}}
\def\hbP{\hat{\Bbb{P}}}
\def\hbQ{\hat{\Bbb{Q}}}
\def\hbH{\hat{\Bbb{H}}}
\def\hbE{\hat{\Bbb{E}}}
\def\hbF{\hat{\Bbb{F}}}
\def\hbG{\hat{\Bbb{G}}}
\def\BbM{{\Bbb{M}}}
\def\BbN{{\Bbb{N}}}
\def\BbP{{\Bbb{P}}}
\def\BbQ{{\Bbb{Q}}}
\def\BbF{{\Bbb{F}}}
\def\BbA{{\Bbb{A}}}
\def\BbL{{\Bbb{L}}}
\def\BbZ{{\Bbb{Z}}}
\def\BbB{{\Bbb{B}}}
\def\BbC{{\Bbb{C}}}
\def\BbH{{\Bbb{H}}}
\def\BbV{{\Bbb{V}}}
\def\BbE{{\Bbb{E}}}
\def\bm{\bar{m}}
\def\bn{\bar{n}}
\def\bz{\bar{z}}
\def\delslash{\del\!\!\!/}
\def\Dslash{D\!\!\!\!/}
\def\Oslash{\cO\!\!\!\!/}

\begin{document}
\begin{flushright}
\hfill{UPR 917T}\\
\hfill{hep-th/0101049}\\
\hfill{January 2001}\\
\end{flushright}


\begin{center}
{ \large {\bf Non-Perturbative Superpotentials from Membrane 
Instantons in Heterotic M-Theory  }}

\vspace{15pt}
\auth

\vspace{15pt}

{\hoch{1}\upenn}

\vspace{10pt}

{\hoch{2}\ufc}

\vspace{40pt}

\underline{ABSTRACT}
\end{center}

     A formalism for calculating the open supermembrane contribution to the
non-perturbative superpotential of moduli in heterotic M-theory is presented.
This is explicitly applied to the Calabi-Yau $(1,1)$-moduli and the separation
modulus of the end-of-the-world BPS three-branes, whose non-perturbative
superpotential is computed. The role of gauge bundles on the boundaries of
the open supermembranes is discussed in detail, and a topological criterion
presented for the associated superpotential to be non-vanishing.

{\vfill\leftline{}\vfill
\footnoterule

{\footnotesize \hoch{ } Research supported in part by DOE grant
DE-AC02-76ER03071 \vskip -12pt} \vskip 14pt
{\footnotesize  \hoch{\dagger} Research supported in part by DAAD (HPS III) 
\vskip  -12pt}}

\pagebreak
\setcounter{page}{1}

\section{Introduction}

In a series of papers \cite{B34,B33,B32,B31}, it was shown that \HW theory 
\cite{HorWit1,HorWit2}, when
compactified on a Calabi-Yau threefold $CY_3$ with non-vanishing $G$-flux,
leads to a natural brane universe theory of particle physics, called
heterotic M-theory. Heterotic M-theory has a five-dimensional bulk space
consisting of a specific gauging of $N=2$ supergravity coupled to both hyper-
and vector supermultiplets. This space has finite length in the fifth-direction,
terminating in two five-dimensional BPS three-branes (eleven-dimensional
nine-branes wrapped on $CY_3$), one, the observable brane, at one end of the
universe and one, the hidden brane, at the other. By constructing holomorphic
vector bundles on Calabi-Yau threefolds with both trivial and non-trivial 
homotopy, it was demonstrated that both three-family grand unified theories
\cite{B27} and the standard model \cite{B15,B5,B4,B3} can appear on the observable 
brane,
while an appropriate hidden sector arises on the other brane. An important
feature of heterotic M-theory is that the non-trivial vector bundles required
to give realistic particle physics gauge groups and spectra generically
predict, via anomaly cancellation, the existence of additional five-dimensional
BPS three-branes (eleven-dimensional five-branes wrapped on a holomorphic
curve $\CC \subset CY_3$). These branes live in the bulk space and represent
new non-perturbative vacua. They may play an important phenomenological role,
both in particle physics \cite{c10,c32,c39,c40,RS,RS'} and cosmology 
\cite{c41,c42,c43,c44,c45}. One interesting
aspect of these bulk branes, their absorption and expulsion
from the boundary three-branes via ``small instanton'' phase transitions,
was discussed in \cite{B12}.

In this paper, we begin an investigation of the stability and dynamics of
non-perturbative heterotic M-theory vacua. All work referenced above assumed
that the potential energy for the various moduli fields in the low energy
four-dimensional effective theory vanished. This is certainly true for moduli
at the perturbative level. However, it is well known 
\cite{WS10,WS11,WJ,BeckerBS,HarvMoor} that 
non-vanishing contributions to the moduli potential energy can arise from
various aspects of non-perturbative physics. Specifically, in the 
four-dimensional $N=1$ supersymmetric effective theory of heterotic M-theory,
a non-vanishing superpotential for moduli, $W$, can arise from non-perturbative
effects. In this paper, we will initiate our study of this topic by
explicitly computing the contribution to $W$ of open supermembranes stretched
between the observable and hidden BPS boundary three-branes. This superpotential
will depend on the $(1,1)$-moduli $T^I$ which carry information about the
K\"{a}hler metric deformations of $CY_3$ and the separation modulus $R$ of
the two boundary three-branes. Specifically, we will do the following.

In Section 2, we discuss BPS membranes in heterotic M-theory and show that they
must stretch between the boundary three-branes in order to preserve $N=1$
supersymmetry. That is, we must consider ``open'' supermembranes only. In the
next section, we present the $\k$-invariant action for an open supermembrane
coupled to the gauge bundles of the boundary three-branes. Section 4 studies
the BPS conditions for the open membrane when wrapped on a complex curve $\CC$
in $CY_3$. It is shown that curve $\CC$ must be holomorphic. In Section 5, we
perform the dimensional reduction of the membrane theory on the 
fifth-dimensional interval $S^1/\Z_2$ and show that this theory becomes the
heterotic superstring coupled to $E_8\times E_8$ and wrapped on holomorphic
curve $\CC$. The following section is devoted to a careful discussion of the 
moduli of heterotic M-theory and their four-dimensional low energy effective
theory. The formalism for computing a non-perturbative contribution to the
superpotential via instanton contributions to the fermion two-point functions
is presented. In Sections 7 and 8, the fermion two-point functions generated
by open supermembranes stretched between the two boundary three-branes are 
explicitly computed; Section 7 discussing the contributions of the membrane
worldvolume while Section 8 computes the contributions of the membrane
boundaries. Finally, in Section 9, using the formalism presented in Section 6
and the results of Sections 7 and 8, we extract the expression for the
non-perturbative superpotential $W$ generated by open supermembranes. We refer
the reader to equation (\ref{final}) for the final result. Various necessary
technical remarks and formalism are presented in two Appendices A and B.
We want to emphasize that our goal in this paper is to compute the 
superpotential for moduli associated with open supermembranes, a quantity
holomorphic on chiral superfields and independent of the dilaton. Since
it is independent of the dilaton, the result, to lowest order, can be computed
taking the low energy limit where the wrapped supermembrane is replaced by the
heterotic string. We take this approach here. Non-holomorphic quantities, 
such as the K\"{a}hler potential, are more subtle, but we do not compute
them in this paper.

Within this context, it is clear that similar calculations can be carried out
for the open supermembrane contributions to the 1) boundary three-brane---bulk
three-brane (eleven-dimensional nine-brane---five-brane) superpotential and
to the 2) bulk three-brane---bulk three-brane (eleven-dimensional
five-brane---five-brane) superpotential. Perhaps not surprisingly, these
calculations and results are similar to those given in this paper, although
there are some fundamental differences. Due to the length of the present
paper, we will present the results involving at least one bulk brane in 
another publication \cite{LOPR}.

The formalism both in this paper and in \cite{LOPR} relies heavily on the ground
breaking work of Strominger, Becker and Becker \cite{BeckerBS} and Harvey and
Moore \cite{HarvMoor}. Recently, a paper due to Moore, Peradze and Saulina 
\cite{MPS}
appeared whose results overlap substantially with the results in this paper
and in \cite{LOPR}. We acknowledge their work and appreciate their pre-announcement
of our independent study of this subject.

\section{Membranes in Heterotic M-Theory:}

\subsection*{Eleven-Dimensional Supergravity and BPS Membranes:}

As is well-known, $D=11$ $N=1$ supersymmetry consists of a single
supergravity multiplet \cite{CJS}. This multiplet contains as its component fields a
graviton $\hat{g}_{\hat{M}\hat{N}}$, a three-form $\hat{C}_{\hat{M}\hat{N}
\hat{P}}$ and a Majorana gravitino $\hat{\Psi}_{\hat{M}}$. The dynamics of this
supermultiplet is specified by a unique action, whose
bosonic terms are 
\bea
S_{S\! G} &=& -\frac{1}{2\k^2} \int_{M_{11}} d^{11} \hat{x}\, 
\sqrt{-\hat{g}} [ \hat{R} + \frac{1}{24}
\hat{G}_{\hat{M}\hat{N}\hat{O}\hat{P}}
\hat{G}^{\hat{M}\hat{N}\hat{O}\hat{P}} 
\nn \\
& & \ \ \ \ \ \ \ \ \ \ + \frac{\sqrt{2}}{1728}
\hat{\e}^{\hat{M}_{1}\ldots\hat{M}_{11}}
\hat{C}_{\hat{M}_{1}\hat{M}_{2}\hat{M}_{3}}
\hat{G}_{\hat{M}_{4}\ldots\hat{M}_{7}}
\hat{G}_{\hat{M}_{8}\ldots\hat{M}_{11}} ] ,\label{11dSG}
\eea
where $(\hat{x}^{\hat{0}},\ldots,\hat{x}^{\hat{9}},\hat{x}^{\hat{11}})$ are 
the eleven-dimensional coordinates, $\hat{g}=\det \hat{g}_{\hat{M}\hat{N}}$ and
$\hat{G}_{\hat{M}\hat{N}\hat{P}\hat{Q}}=
24\del_{[\hat{M}}\hat{C}_{\hat{N}\hat{P}\hat{Q}]}$is the field strength of 
the three-form $\hat{C}$ defined by $\hat{G}=d\hat{C}$. This action is invariant under the
supersymmetry transformations of the component fields. For our purposes, we
need only specify the supersymmetry variation of the gravitino field 
$\hat{\Psi}_{\hat{M}}$, which is given by
\be
\d_{\hat{\vare}} 
\hat{\Psi}_{\hat{M}} = \hat{D}_{\hat{M}} \hat{\vare} +
\frac{\sqrt{2}}{288}
(\hat{\G}_{\hat{M}}^{\; \hat{N}\hat{P}\hat{Q}\hat{R}} - 8 
\d_{\hat{M}}^{\hat{N}} \hat{\G}^{\hat{P}\hat{Q}\hat{R}} ) \hat{\vare}
\hat{G}_{\hat{N}\hat{P}\hat{Q}\hat{R}} + \cdots , \label{11dSusy}
\ee
where $\hat{\vare}$ is the Majorana supersymmetry parameter and the dots denote
terms that involve the fermion fields of the theory.
The 11-dimensional spacetime Dirac matrices $\hat{\G}_{\hat{M}}$ satisfy
$\{\hat{\G}_{\hat{M}},\hat{\G}_{\hat{N}}\}=2\hat{g}_{\hat{M}\hat{N}}$.
Throughout this paper, we will follow convention and refer to the theory of
$D=11$, $N=1$ supergravity as $M$-theory, although the complete $M$-theory 
must really include, presently unknown, higher order physics.

It is also well-known that there is a $2+1$-dimensional membrane solution of 
the $M$-theory equations of motion that preserves one-half of the
supersymmetries \cite{DuffStelle}. This ``electrically charged'' solution is 
equivalent, upon double dimensional reduction on $S^1$, 
to the elementary string solution of the ten-dimensional type $I\! IA$ 
supergravity equations of motion. This BPS membrane solution
is of the form
\bea
d \hat{s}^2 &=&  e^{2A} d \hat{z}^{\hi} d \hat{z}^{\hj} \,
\hat{\h}_{\hi \hj} + e^{2B} d \hat{y}^{\hat{m}} d \hat{y}^{\hat{n}} \,
\d_{\hat{m}\hat{n}}, \nn \\
\hat{C}_{\hi \hj \hat{k}} &=& - \frac{1}{6\sqrt{2}}
\hat{\vare }_{\hi \hj \hat{k}} \, e^{C}, \label{11dMsol}
\eea
where $\hat{z}^{\hi}$ are the three coordinates oriented in the direction 
of the membrane (with $\hi = \hat{0},\hat{1},\hat{2}$)
and $\hat{y}^{\hat{m}}$ ($\hat{m} = \hat{3},\ldots,\hat{10}$)
are the eight coordinates transverse to the membrane. Note that these
coordinates do not necessarily coincide with $\hat{x}^{\hat{M}}$.
Generically, there is a sign ambiguity on
the right-hand side of the second equation in (\ref{11dMsol}). The choice of
this sign defines an orientation of the preserved supersymmetry on the 
membrane. In this paper,
we have arbitrarily chosen the $-$ sign. That corresponds to positive
chiral supersymmetry from the normal spin bundle point of view. This choice
of sign is for specificity only, our conclusions being independent of it. 
The functions $A,B,C$ depend on the transverse 
coordinates only. Equations (\ref{11dMsol}) represent a solution of 
eleven-dimensional supergravity when 
\be
A = -2B = C/3, \ \ \ \ \ \mbox{and} \ \ \ \ \ 
e^{-C} = 1 + \frac{1}{\hat{r}^6},
\ee
where
\be
\hat{r} \equiv \sqrt{(\hat{y}^{\hat{m}} - \hat{y}^{\hat{m}}_{0})
 (\hat{y}^{\hat{n}} - \hat{y}^{\hat{n}}_{0}) \d_{\hat{m}\hat{n}} }
\ee
and $\hat{y}^{\hat{m}}_{0}$ are constants.
This solution describes a membrane located at $(\hat{y}^{\hat{3}}_{0},
\ldots,\hat{y}^{\hat{10}}_{0})$.

That this solution preserves one-half of the supersymmetry can be seen as 
follows. The
supersymmetry transformation of the gravitino is given in (\ref{11dSusy}). Now
make the three-eight split
\be
\hat{\G}_{\hat{A}} = (\hat{\t}_{\hat{a}'} \otimes \tilde{\hat{\g}},
1 \otimes \hat{\g}_{\hat{a}} ),
\ee
where $\hat{\t}_{\hat{a}'}$ and $\hat{\g}_{\hat{a}}$ are the three- and
eight-dimensional Dirac matrices, respectively, with flat indices 
$\hat{A}=\hat{0},\ldots,\hat{10}$,
$\hat{a}'=\hat{0},\hat{1},\hat{2}$ and
$\hat{a}=\hat{3},\ldots,\hat{10}$, and 
\be
\tilde{\hat{\g}}=\prod_{\hat{a}=\hat{3}}^{\hat{10}} \hat{\g}_{\hat{a}}. 
\ee 
Then the supersymmetry variation (\ref{11dSusy}) vanishes
for spinor parameters $\hat{\vare}$ of the form
\be
\hat{\vare} = \hat{\l}_0 \otimes \hat{\n}_0 \, e^{C/6} ,
\ee
where $\hat{\l}_0$ and $\hat{\n}_0$ are constant three- and eight-dimensional
spinors, and $\hat{\n}_0$ satisfies the chirality condition
\be
\frac{1}{2}(1 - \tilde{\hat{\g}} ) \hat{\n}_0 = 0. \label{11dChir}
\ee
The minus sign in (\ref{11dChir}), which determines the chirality of the 
preserved supersymmetry, is correlated to the sign arbitrarily chosen in 
the membrane configuration (\ref{11dMsol}).

This solution solves the eleven-dimensional supergravity
equations of motion everywhere except at the singularity
$\hat{r}=0$. This implies that it is necessary to include delta function
source terms in the supergravity equations of motion \cite{DuffStelle}.
The source terms that must be added to the pure supergravity action are 
precisely the supermembrane action (\cite{BST})
\be
S_{S\!M} =- T_M \int_{\Si} d^3 \hat{\s} ( 
\sqrt{-\det \hat{g}_{\hi\hj}} - \frac{1}{6} 
\hat{\vare}^{\hi\hj \hat{k}} \hat{\Pi}_{\hi}^{\, \hbA}
\hat{\Pi}_{\hj}^{\, \hbB} \hat{\Pi}_{\hat{k}}^{\, \hbC} 
\hbC_{\hbC\hbB\hbA} ), \label{SMaction}
\ee
where 
\be
T_M = (2 \pi^2 / \k^2)^{1/3}
\ee
is the membrane tension of mass dimension three,
\be
\hat{g}_{\hi\hj} = \hat{\Pi}_{\hi}^{\, \hat{A}} \hat{\Pi}_{\hj }^{\,
\hat{B}}
\h_{\hat{A}\hat{B}}, \ \ \ \ \ \ \ 
\hat{\Pi}_{\hi}^{\, \hbA} = \del_{\hi}\hbZ^{\hbM} 
\hbE_{\hbM}^{\; \hbA}, \label{supermetric}
\ee
and $\hat{\s}^{\hat{0}},\hat{\s}^{\hat{1}},\hat{\s}^{\hat{2}}$ are the 
worldvolume coordinates of the membrane
which are, generically, independent of any target space coordinates.
This action represents the superembedding $\hbZ : 
\Si^{3|0} \ra M^{11|32}$, 
whose bosonic and fermionic component
fields are the background coordinates, separated as
\be
\hbZ^{\hbM}(\hat{\s}) = (\hat{X}^{\hat{M}}(\hat{\s}) , 
\hat{\T}^{\hat{\m}} (\hat{\s}) ), \label{supercoord}
\ee
respectively.
The action is a sigma-model since the super-elfbeins 
$\hbE_{\hbM}^{\; \hbA}$ and the super-three-form
$\hbC_{\hbC\hbB\hbA}$ both depend on the superfields 
$\hbZ^{\hbM}$. The super-elfbeins have, as their first bosonic and fermionic 
component in the $\hat{\T}$ expansion, the bosonic elfbeins 
$\hat{E}_{\hat{M}}^{\, \hat{A}}$
and the gravitino $\hat{\Psi}_{\hat{M}}^{\; \hat{\a}}$
respectively, while the super-three-form has the bosonic three-form from 
eleven-dimensional supergravity as its leading field component.
The superfields $\hbE_{\hbM}^{\; \hbA}$ and $\hbC_{\hbC\hbB\hbA}$
represent the background into which the supermembrane is embedded and, 
therefore, must satisfy the eleven-dimensional supergravity equations of
motion \cite{cremmferr}. Generically, there is a sign ambiguity in the 
second term 
of (\ref{SMaction}). Here we choose the sign that is consistent with our 
choice of sign in Eq.(\ref{11dMsol}). 

The fact that membrane configuration (\ref{11dMsol}) is a solution of
$M$-theory which preserves one-half of the supersymmetries
translates, when speaking in supermembrane worldvolume language, into the fact
that the action (\ref{SMaction}) exhibits a local fermionic invariance,
$\kappa$-invariance, that is used to gauge away half of the fermionic 
degrees of freedom.
Specifically, the supermembrane action is invariant under the local 
fermionic symmetries
\be
\d_{\hat{\k}} \hat{\T} = 2 \hat{P}_+ \hat{\k} + \cdots, \ \ \ \ \ 
\d_{\hat{\k}} \hat{X}^{\hat{M}} = 2 \bar{\hat{\T}} 
\hat{\G}^{\hat{M}} \hat{P}_+ \hat{\k} + \cdots , \label{kappa}
\ee
where $\hat{\k} (\hat{\s})$ is an eleven-dimensional local spinor
parameter and 
$\hat{P}_{\pm}$ are the projection operators
\be
\hat{P}_{\pm} \equiv \frac{1}{2} (1 \pm \frac{1}{6\sqrt{-\det
\hat{g}_{\hi\hj}}} 
\hat{\vare}^{\hi\hj \hat{k}} \hat{\Pi}_{\hi}^{\, \hat{A}} 
\hat{\Pi}_{\hj }^{\, \hat{B}} \hat{\Pi}_{\hat{k}}^{\, \hat{C}} 
\hat{\G}_{\hat{A}\hat{B}\hat{C}} ), \label{projop}
\ee
obeying
\be
\hat{P}_{\pm}^2 = \hat{P}_{\pm} , \ \ \ \ \hat{P}_+ \hat{P}_- = 0 ,
\ \ \ \ \hat{P}_+ + \hat{P}_- = 1.
\ee
It follows from the first equation in (\ref{kappa}) that the $\hat{P}_+\hat{\T}$
component of spinor $\hat{\T}$ can be transformed away by a $\k$-transformation.
Note that (\ref{kappa}) includes only the leading order terms in $\hat{\T}$,
which is all that is required to discuss the supersymmetry properties of the
membrane. The theory is not, in general, invariant under global supersymmetry 
transformations
\be
\d_{\hat{\vare}} \hat{\T} = \hat{\vare}, \ \ \ \ \ 
\d_{\hat{\vare}} \hat{X}^{\hat{M}} = \bar{\hat{\vare}} 
\hat{\G}^{\hat{M}} \hat{\T},        \label{glfermTransf}
\ee
where $\hat{\vare}$ is an eleven-dimensional spinor independent of
$\hat{\s}$.
For example, a general bosonic configuration $\hat{X}(\hat{\s})$ breaks
all global supersymmetries generated by $\hat{\vare}$. However,
one-half of the supersymmetries will remain unbroken if and only if
(\ref{glfermTransf}) can be compensated for by a $\k$-transformation with a 
suitable parameter $\hat{\k}(\hat{\s})$. That is
\bea
\d \hat{\T} &=& \d_{\hat{\vare}} \hat{\T} + 
\d_{\hat{\k}} \hat{\T} \nn \\
&=& \hat{\vare} + 2 \hat{P}_+ \hat{\k} (\hat{\s}) = 0. \label{kappaTheta}
\eea
In order for this to be satisfied, a necessary condition is that
\be
\hat{P}_- \hat{\vare} = 0. \label{11dBPS} 
\ee
Of course, the sign choice of
$\hat{P}_+$ in (\ref{kappa}) is correlated to the sign of the second term
in the supermembrane action (\ref{SMaction}), the sign which fixes 
the membrane supersymmetry chirality \cite{BST}. 
Had we chosen the opposite chirality, the 
symbols $\hat{P}_+$ and $\hat{P}_-$ would be interchanged in 
the present discussion. 

\subsection*{Membranes in \HW Theory:}

When M-theory is compactified on $S^1/\Z_2$, it describes the low energy limit
of the strongly coupled heterotic string theory \cite{HorWit1,HorWit2}. 
We choose $\hat{x}^{\hat{11}}$ as the orbifold direction and parametrize $S^1$
by $\hat{x}^{\hat{11}} \in [-\pi\r ,\pi\r]$ with the endpoints identified.
The $\Z_2$ symmetry acts by 
further identifying any point $\hat{x}^{\hat{11}}$ with 
$-\hat{x}^{\hat{11}}$ and, therefore, gives rise to two ten-dimensional 
fixed hyperplanes at $\hat{x}^{\hat{11}}=0$ and $\hat{x}^{\hat{11}}=\pi \r$. 
Since, at each 
$\Z_2$ hyperplane, only the field components that are even under the $\Z_2$
action can survive, the eleven-dimensional supergravity in the bulk 
space is projected
into $N=1$ ten-dimensional chiral supergravity on each boundary. Furthermore,
cancellation of the chiral anomaly in this theory requires the existence of 
an $N=1$,
$E_{8}$ super-Yang-Mills multiplet on each fixed hyperplane 
\cite{HorWit1,HorWit2}.
Therefore, the effective action
for M-theory on $S^1 / \Z_2$ describes the coupling of two ten-dimensional
$E_8$ super-Yang-Mills theories, one on each hyperplane, to eleven-dimensional 
supergravity in the bulk space. 
The bosonic part of the \HW action is given by
\be
S_{H\! W}=S_{S\! G}+S_{Y\! M},
\ee
where $S_{S\! G}$ is the eleven-dimensional supergravity bulk
action (\ref{11dSG})
and $S_{Y\! M}$ describes the two $E_8$ Yang-Mills theories on the 
orbifold planes
\be
S_{Y\! M} =  -\frac{1}{8 \p \k^2} (\frac{\k}{4 \p})^{2/3} \sum_{n=1,2} 
\int_{M_{10}^{(n)}} d^{10} \, x \sqrt{-g} [ \mbox{tr} (F^{(n)})^2 -
\frac{1}{2} 
\mbox{tr} R^2 ]  \label{BoundAction}
\ee
Here $F^{(n)}_{MN}$, with $M,N=0,1,\ldots,9$ and $n=1,2$, are the two $E_8$ 
gauge field-strengths where 
$F^{(n)}_{MN}=F^{(n)a}_{MN} T^a$ ($T^a$ being $E_8$ generators, $a=1,
\ldots,248$).
$R$ is the eleven-dimensional curvature two-form restricted to the
orbifold 
planes $M_{10}^{(n)}$. 
The supersymmetry transformations of the fermionic fields are
given by (\ref{11dSusy}) for the gravitino and
\be
\d_{\hat{\vare}} \vart^{(n)a} = F^{(n)a}_{AB} \G^{AB} \hat{\vare} + 
\cdots  \label{gaugino}
\ee
for the fixed hyperplane gauginos $\vart^{(n)a}$, where $A,B=0,1,\ldots,9$
are flat ten-dimensional indices.

In order to cancel all chiral anomalies on the hyperplanes, the action 
$S_{H\! W}$ has to be supplemented by the modified Bianchi 
identity\footnote{The normalization of $\hat{G}$ adopted here differs from 
\cite{HorWit1} by a factor of $\sqrt{2}$ but it agrees with \cite{Ceder},
in which one considers, as we will do in this paper, the superfield
version of the Bianchi identities.}
\be
(d \hat{G})_{\hat{11}MNPQ} = - \frac{1}{4 \pi} 
(\frac{\k}{4 \pi})^{2/3} \{ J^{(1)} \d (\hat{x}^{\hat{11}}) +
J^{(2)} \d (\hat{x}^{\hat{11}} - \pi \r) \}_{MNPQ} , \label{modBianchi}
\ee
where
\be
J^{(n)} = \mbox{tr} F^{(n)} \wedge F^{(n)} - \frac{1}{2} \mbox{tr} R \wedge R,
\ee
for $n=1,2$. 
The solutions to the equations of motion resulting from action
$S_{H\! W}$ must respect the $\Z_2$ orbifold symmetry. Under $\Z_2$,
the bosonic fields in the bulk behave as
\be
\ba{rl}
\hat{g}_{MN} (\hat{x}^{\hat{11}}) = \hat{g}_{MN} (-\hat{x}^{\hat{11}}),
& \ \ \ \ \hat{C}_{MNP} (\hat{x}^{\hat{11}}) = - \hat{C}_{MNP} 
(-\hat{x}^{\hat{11}}), \\
\hat{g}_{M\hat{11}} (\hat{x}^{\hat{11}}) = - \hat{g}_{M\hat{11}} 
(-\hat{x}^{\hat{11}}), & \ \ \ \ \hat{C}_{\hat{11}MN} (\hat{x}^{\hat{11}})
= 
\hat{C}_{\hat{11}MN} (-\hat{x}^{\hat{11}}), \\  \hat{g}_{\hat{11}\hat{11}} 
(\hat{x}^{\hat{11}}) = \hat{g}_{\hat{11}\hat{11}} (-\hat{x}^{\hat{11}}). 
& \ea \label{bosuz2}
\ee
while the gravitino transforms as
\be
\hat{\Psi}_{M} (\hat{x}^{\hat{11}}) = \hat{\G}_{\hat{11}}
\hat{\Psi}_{M} (-\hat{x}^{\hat{11}}), \ \ \ \ \ \ \ 
\hat{\Psi}_{\hat{11}} (\hat{x}^{\hat{11}}) = -
\hat{\G}_{\hat{11}} \hat{\Psi}_{\hat{11}} (-\hat{x}^{\hat{11}}).
\label{psiuz2}
\ee
where $\hat{\G}_{\hat{11}}=\hat{\G}_{\hat{0}}\hat{\G}_{\hat{1}}
\cdots\hat{\G}_{\hat{9}}$. In order for the supersymmetry transformations 
of the gravitino to be consistent with the $\Z_2$ symmetry, the 
eleven-dimensional Majorana spinor in (\ref{11dSusy}) $\hat{\vare}$ must satisfy
\be
\hat{\vare} (\hat{x}^{\hat{11}}) = \hat{\G}_{\hat{11}}
\hat{\vare} (-\hat{x}^{\hat{11}}). \label{epsuz2}
\ee
This equation does not restrict the number of independent components of the
spinor fields $\hat{\vare}$ at any point in the bulk space. However, at 
each of the $\Z_2$ hyperplanes, constraint (\ref{epsuz2}) becomes the 
ten-dimensional chirality condition
\be
\frac{1}{2} (1 - \hat{\G}_{\hat{11}}) \hat{\vare} = 0, \ \ \ \ \ \  
\mbox{at} \ \ \ \hat{x}^{\hat{11}} = 0, \pi \r. \label{z2chiral}
\ee
This leads to the correct amount of supersymmetry, $N=1$, on each of the 
ten-dimensional orbifold fixed planes. Generically, there is a sign 
ambiguity in equations (\ref{psiuz2}) and (\ref{epsuz2}). The choice
made here coincides with \cite{HorWit1}. This choice is consistent with the
previous choice of supersymmetry orientation of the 
membrane.

Membrane solutions were explicitly 
constructed for \HW theory in \cite{LalLukOvr}. There, the membrane
solution of eleven-dimensional supergravity was shown to satisfy, when
appropriately modified and the boundary gauge fields are turned off, 
the equations of motion of the theory subject to the $\Z_2$ constraints.

There are two different ways to orient the membrane with
respect to the orbifold direction, that is,  $\hat{x}^{\hat{11}}$ can either 
be a transverse coordinate or a coordinate oriented in the direction of
the membrane. In the first case, the membrane 
is parallel to the hyperplanes. In the second case, it
extends between the two hyperplanes and intersects each of
them along a $1+1$-dimensional string. This latter configuration is 
sometimes referred to as an open supermembrane.

Let us consider the question of which type of supersymmetry each of these
two membrane configurations preserve. Note that each such membrane is, by
construction, a supersymmetry-preserving solution of supergravity. The 
associated singular worldvolume theory is, according to the previous section, 
a supersymmetric theory in the sense that is has $\k$-invariance that can 
gauge away only half of the fermions.\footnote{We will discuss the 
$\k$-invariance of the open supermembrane worldvolume theory in the next 
section.}
Therefore, each of the two membrane orientations
discussed above preserves one-half of the supersymmetries. More interesting is 
the question as to whether the 
particular supersymmetries that each of the two configurations preserves 
can be made consistent with the supersymmetry, defined in 
Eq.(\ref{z2chiral}), that is imposed on the orbifold fixed hyperplanes. 

We have seen in (\ref{11dBPS}) that, in order for supersymmetry to be
preserved, the global supersymmetry parameter $\hat{\vare}$ of the membrane
worldvolume theory must satisfy
$\hat{P}_- \hat{\vare}=0$, where $\hat{P}_-$ is given in (\ref{projop}).
Consider first a configuration in which the membrane is oriented parallel
to the ten-dimensional hyperplanes. Choose the fields $\hat{X}$ and $\hat{\T}$ 
such that
\bea
\hat{X}^{\hat{0}} &=& \hat{\s}^{\hat{0}}, \ \ \ \ \ \ \ 
\hat{X}^{\hat{1}} = \hat{\s}^{\hat{1}}, \ \ \ \ \ \ \ 
\hat{X}^{\hat{2}} = \hat{\s}^{\hat{2}}, \nn \\
\hat{X}^{\hat{m}} &=& 0, \ \ (\hat{m}=3,\ldots,9,11) \ \ \ \ \ \ \ 
\hat{\T} = 0.
\eea
Then $\hat{P}_- \hat{\vare} = 0$ simplifies to
\be
\hat{P}_- \hat{\vare} = \frac{1}{2} (1 - \hat{\G}_{\hat{0}}
\hat{\G}_{\hat{1}}\hat{\G}_{\hat{2}}) \hat{\vare} = 0. \label{badconf}
\ee
Because the membrane is embedded in a ten-dimensional space perpendicular to 
the orbifold direction, we need only consider eleven-dimensional spinors on the
membrane worldvolume that can be decomposed into linearly independent chiral spinors 
\be
\hat{\vare} = \hat{\vare}_+ + \hat{\vare}_-,
\ee
where
\be
\hat{\vare}_{\pm}=\frac{1}{2}(1\pm\hat{\G}_{\hat{11}})\hat{\vare} .
\label{10dimchir}
\ee
Then (\ref{badconf}) becomes
\be
\frac{1}{2} (1 - \hat{\G}_{\hat{0}} \hat{\G}_{\hat{1}}\hat{\G}_{\hat{2}}) 
\hat{\vare}_+ = 0, \ \ \ \ \ \ \ \ \frac{1}{2} (1 - \hat{\G}_{\hat{0}}
\hat{\G}_{\hat{1}}\hat{\G}_{\hat{2}}) \hat{\vare}_- = 0 .
\ee
Now multiply these expressions on the left by $\hat{\G}_{\hat{11}}$ and
use the chirality properties of $\hat{\vare}_{\pm}$ defined by 
(\ref{10dimchir}). It follows that
\be
\frac{1}{2} (1 + \hat{\G}_{\hat{0}} \hat{\G}_{\hat{1}}\hat{\G}_{\hat{2}}) 
\hat{\vare}_+ = 0, \ \ \ \ \ \ \ \ \frac{1}{2} (1 + \hat{\G}_{\hat{0}}
\hat{\G}_{\hat{1}}\hat{\G}_{\hat{2}}) \hat{\vare}_- = 0 .
\ee
Combining the two sets of equations gives $\hat{\vare}_{\pm}=0$. 
Therefore, 
even though the membrane preserves one-half of the supersymmetries, they
do not coincide with the supersymmetries preserved on the boundaries.

Next, consider a configuration in which the membrane is oriented
perpendicular to the ten-dimensional hyperplanes. We choose the fields such that
\bea
\hat{X}^{\hat{0}} &=& \hat{\s}^{\hat{0}}, \ \ \ \ \ \ \ 
\hat{X}^{\hat{1}} = \hat{\s}^{\hat{1}}, \ \ \ \ \ \ \ 
\hat{X}^{\hat{11}} = \hat{\s}^{\hat{2}}, \nn \\
\hat{X}^{\hat{m}} &=& 0, \ \ \ (\hat{m}=2,3,\ldots,9) \ \ \ \ \ \ \ 
\hat{\T} = 0, \label{rightconf}
\eea
so that $\hat{P}_- \hat{\vare} = 0$ now becomes
\be
\hat{P}_- \hat{\vare} = \frac{1}{2} (1 - \hat{\G}_{\hat{0}}
\hat{\G}_{\hat{1}}\hat{\G}_{\hat{11}})
\hat{\vare} = 0. \label{goodconf}
\ee
This is as far as one can go in the bulk space. However,
on the orbifold boundary planes, (\ref{z2chiral}) can be substituted in
(\ref{goodconf}) to give
\be
\frac{1}{2} (1 - \hat{\G}_{\hat{0}\hat{1}})
\hat{\vare} = 0, \ \ \ \ \  \mbox{at} \ \ \ \hat{x}^{\hat{11}} = 0, \pi \r
.
\label{boundBPS} 
\ee
This expression implies that the eleven-dimensional Majorana 
spinor $\hat{\vare}$,
when restricted to the $1+1$-dimensional boundary strings (thereafter 
denoted by $\vare$), is a non-vanishing Majorana-Weyl spinor, as it should 
be.\footnote{When we switch to Euclidean space later in this paper, 
we must regard $\hat{\vare}$ as an eleven-dimensional Dirac spinor and
$\vare$ as a ten-dimensional Weyl spinor, since in these dimensions one
cannot impose the Majorana condition.}
We thus see that this configuration preserves one-half of the  
supersymmetries on the $\Z_2$ hyperplanes.

Therefore, we conclude that a configuration in which the supermembrane
is oriented parallel to the orbifold hyperplanes breaks all
supersymmetries. On the other hand, the configuration
for the open supermembrane is such that the hyperplane and membrane 
supersymmetries are compatible.

\section{$\k$-Invariant Action for Open Membranes:}

We have shown that for a supermembrane  
to preserve supersymmetries consistent with the boundary fixed planes,
the membrane must be open,
that is, stretched between the two $\Z_2$ hyperplanes.
In this section, we want to find the action associated with such a membrane.
Action  
(\ref{SMaction}) is a good starting point. However, it is not obvious that
it will correspond to the desired configuration, even 
in the bulk space. For this to be the case, one needs to ask whether this 
action respects the $\Z_2$ 
symmetry of \HW theory. The answer was provided in \cite{LalLukOvr}, where 
it was concluded that, for an appropriate extension of the $\Z_2$ symmetry to
the worldvolume coordinates and similar constraints for the worldvolume metric,
the open supermembrane equations of motion are indeed $\Z_2$ covariant. 
Therefore,
we can retain action (\ref{SMaction}). Does it suffice, however, to completely
describe the open membrane configuration? Note that the intersection of an
open membrane with each orbifold fixed plane is a $1+1$-dimensional string 
embedded in the ten-dimensional boundary. We denote by $\s^i$, $i=1,2$, the
worldsheet coordinates of these strings.
Intuitively, one expects extra fields, which we generically denote by 
$\phi(\s)$, to appear on each boundary string in addition to the bulk 
fields $\hbZ^{\hbM}(\hat{\s})$. These would naturally couple to the pullback 
onto each boundary string of the
background $E_8$ super-gauge fields $\BbA_{\BbM}$.
As we will see in this section, new supermembrane
fields are indeed required and form a chiral Wess-Zumino-Witten multiplet for 
each $E_8$ gauge group.\footnote{This section follows closely the original proof
in \cite{Ceder}.}

As discussed previously, the supergravity theory of the
background fields exhibits both gauge and gravitational anomalies that can
only be cancelled by modifying the Bianchi identity as in (\ref{modBianchi}). 
Integrating (\ref{modBianchi}) along the 
$\hat{x}^{\hat{11}}$ direction, and promoting the result to superspace, we find
for the $n$-th boundary plane that
\be
\hbG_{\BbM\BbN\BbP\BbQ} \mid_{M^{(n)}_{10}} = - \frac{1}{8 \pi T_M} 
( \mbox{tr} \BbF^{(n)} \wedge \BbF^{(n)} )_{\BbM\BbN\BbP\BbQ} , 
\label{superBianchi}
\ee
where $\BbF^{(n)}$ is the super-field-strength of the fields $\BbA^{(n)}$.
Note that we have dropped the curvature term in the modified Bianchi
identity, since it is a higher-order derivative term and it does not
contribute to the one-loop calculation of the superpotential in this paper.
As a consequence, we are allowed to use the eleven-dimensional supergravity
in the bulk space.
The reason for expressing the integrated Bianchi identity in superspace is
to make it compatible with the bulk part of
supermembrane action (\ref{SMaction}), which is written in terms of 
the pullbacks
of superfields $\hbE_{\hbM}^{\; \hbA}$ and $\hbC_{\hbC\hbB\hbA}$ onto the 
worldvolume. Noting that, locally, $\hbG = d \hbC$, it follows from 
(\ref{superBianchi}) that on the $n$-th boundary plane
\be
\hbC_{\BbM\BbN\BbP} \mid_{M^{(n)}_{10}} =
 - \frac{1}{8 \pi T_M} \Omega_{\BbM\BbN\BbP} (\BbA^{(n)}), \label{CS}
\ee
where
\be
\Omega_{\BbM\BbN\BbP} (\BbA^{(n)}) = 3 !
\left( \mbox{tr} (\BbA^{(n)} \wedge d \BbA^{(n)} ) 
+ \frac{2}{3} \mbox{tr} (\BbA^{(n)} \wedge \BbA^{(n)} \wedge \BbA^{(n)})
\right)_{\BbM\BbN\BbP} \label{CS2}
\ee
is the Chern-Simons three-form of the super-one-form $\BbA^{(n)}$.

Note that each $\BbA^{(n)}$ is a super-gauge-potential and, as such,
transforms under super-gauge transformations as 
\be
\d_{\BbL} \BbA_{\BbM}^a = \del_{\BbM} \BbL^a + f^{abc} 
\BbA_{\BbM}^b \BbL^c, \label{gaugtransf}
\ee
with $a,b,c=1,\ldots,248$. Note that for simplicity, here and elsewhere where it
is inessential, we drop the superscript $(n)$ indicating the boundaries. 
If we define the pullback of $\BbA$ as
\be
\BbA_i \equiv \del_i \BbZ^{\BbM} \BbA_{\BbM},
\ee
the gauge transformation in superspace (\ref{gaugtransf}) induces a gauge
transformation on the string worldsheet, which acts on the 
pullback of $\BbA$ as
\be
\d_{\BbL} \BbA_i^a = (D_i \BbL)^a = \del_i \BbL^a + f^{abc} 
\BbA_i^b \BbL^c,
\ee
where $\BbL = \BbL(\BbZ^{\BbM}(\s))$. It follows from (\ref{CS}), 
(\ref{CS2}) and (\ref{gaugtransf}) that, on each boundary fixed plane,
\be
\d_{\BbL} \hbC_{\BbM\BbN\BbP} = - \frac{3}{4 \pi T_M} \left[
\d_{\BbL} \left( \mbox{tr} (\BbA \wedge d \BbA) 
+ \frac{2}{3} \mbox{tr} (\BbA \wedge \BbA \wedge \BbA) 
\right)_{\BbM\BbN\BbP} \right] = - \frac{3}{4 \pi T_M} 
\mbox{tr} ( \del_{[\BbM} \BbL \del_{\BbN} \BbA_{\BbP ]} ) .
\ee
Now consider the variation of the supermembrane action (\ref{SMaction}) under 
a super-gauge transformation. Clearly, a non-zero variation arises from the
second term in (\ref{SMaction})
\bea
\d_{\BbL} S_{S\!M}  &=& \frac{T_M}{6} \int_{\Si} d^3 \hat{\s} \; 
\hat{\vare}^{\hi\hj \hat{k}} \del_{\hi}\hbZ^{\hbM}
\del_{\hj}\hbZ^{\hbN} \del_{\hat{k}}\hbZ^{\hbP} 
\d_{\BbL} \hbC_{\hbP\hbN\hbM} \nn \\
 &=& \frac{1}{8 \pi} \int_{\del \Si} d^2 \s \;
\vare^{ij} \del_i \BbZ^{\BbM} \del_j \BbZ^{\BbN}
\mbox{tr} ( \BbL \del_{\BbN} \BbA_{\BbM} ),  \label{SuG}
\eea
where $\del \Si$ is the sum over the two boundary strings $\sum_{n=1,2}
\del \Si^{(n)}$, and we have integrated by parts. Therefore, action
(\ref{SMaction}) is not invariant under gauge transformations. This symmetry
is violated precisely at the boundary planes. It follows that to restore
gauge invariance, one must add appropriate boundary terms
to the supermembrane action.

Before doing that, however, let us consider the transformation of the action
$S_{S\!M} $ under a $\k$-transformation, taking into account 
the boundary expression (\ref{CS}). Note that the $\k$-transformation
acts on the super-three-form $\hbC$ as
\be
\d_{\hat{\k}} \hbC = {\cal L}_{\hat{\k}} \hbC
= i_{\hat{\k}}\, d \hbC  + (d i_{\hat{\k}})\, \hbC\, , \label{Liederiv}
\ee
where ${\cal L}_{\hat{\k}}$ is the Lie derivative in the $\hat{\k}$-direction
and the operator $i_{\hat{\k}}$ is defined, for any super-$l$-form $\hbH$, as
\bea
i_{\hat{\k}} \hbH &=& \frac{1}{l!}\hbH_{\hbM_1 \cdots \hbM_l} i_{\hat{\k}} ( d 
\hbZ^{\hbM_l} \wedge \cdots \wedge d \hbZ^{\hbM_1} ) \nn \\
&=& \frac{1}{(l-1)!}\hbH_{\hbM_1 \cdots \hbM_{l-1} \hat{\m}}\, (\hat{P}_+ 
\hat{\k}^{\hat{\m}} )( d \hbZ^{\hbM_{l-1}} \wedge \cdots \wedge d \hbZ^{\hbM_1}
). \eea
Importantly, we use the positive projection $\hat{P}_+$ of $\hat{\k}$, as 
defined in (\ref{kappa}), in order to remain consistent
with the previous choices of supersymmetry orientation. Varying action 
(\ref{SMaction}) under (\ref{Liederiv}) and under the full $\k$-variations
of $\hbZ$,
we observe that $\k$-symmetry is also violated at the boundaries
\bea
\d_{\hat{\k}} S_{S\!M}  &=& - \frac{1}{6} T_M \int_{\del \Si} d^2 \s \;
\vare^{ij} \del_i \BbZ^{\BbM} \del_j \BbZ^{\BbN}
\BbC_{\BbN \BbM \hat{\m}} \hat{P}_+ \hat{\k}^{\hat{\m}} \nn \\ 
&=& \frac{1}{48 \pi} 
\int_{\del \Si} d^2 \s \;
\vare^{ij} \del_i \BbZ^{\BbM} \del_j \BbZ^{\BbN}
\Omega_{\BbN \BbM \hat{\m}} (\BbA) \hat{P}_+ \hat{\k}^{\hat{\m}}.
\eea
In deriving this expression we have used the eleven-dimensional supergravity
constraints.
It proves convenient to consider, instead of this $\k$-transformation, the
modified $\k$-transformation
\be
\D_{\hat{\k}} = \d_{\hat{\k}} - \d_{\BbL_{\hat{\k}}}, \label{gkappa}
\ee
where $\d_{\BbL_{\hat{\k}}}$ is a super-gauge transformation with the special
gauge parameter
\be
\BbL_{\hat{\k}} = i_{\hat{\k}} \BbA = 2 \BbA_{\hat{\m}} \hat{P}_+ 
\hat{\k}^{\hat{\m}}. 
\ee
Under this transformation, the supermembrane action behaves as
\be
\D_{\hat{\k}} S_{S\!M}  = \frac{1}{8 \pi} \int_{\del \Si} d^2 \s \;
\vare^{ij} \del_i \BbZ^{\BbM} \del_j \BbZ^{\BbN}
\BbF_{\BbN \hat{\m}} \hat{P}_+ \hat{\k}^{\hat{\m}} \BbA_{\BbM}. \label{SuK}
\ee
It is also useful to note that the pullback of the boundary background field 
$\BbA$ transforms as
\be
\D_{\hat{\k}} \BbA_i = 2 \del_i \BbZ^{\BbM} \BbF_{\BbM \hat{\m}}
\hat{P}_+ \hat{\k}^{\hat{\m}}
\ee
under this modified $\k$-transformation, where we have used the fact that
\be
\d_{\hat{\k}} \BbA = {\cal L}_{\hat{\k}} \BbA
\ee
is the $\k$-transformation of $\BbA$, just as in (\ref{Liederiv}).

It was shown in \cite{Ceder} that the gauge and modified $\k$ anomalies
can be cancelled if the supermembrane action is augmented to include a
chiral level one Wess-Zumino-Witten model on each boundary string 
of the membrane. The fields thus introduced will couple to the pullback of the 
background fields $\BbA$ at each boundary.

On each boundary string, the new fields can be written as
\be
g(\s)=e^{\phi^a (\s) T^a }, \label{e8}
\ee
where $T^a$ are the generators of $E_8$ (with $a=1,\ldots,248$) 
and $\phi^a (\s)$ are scalar fields that transform in the adjoint 
representation, and parametrize the group manifold, of $E_8$.
Note that $g$ is a field living on the worldsheet of the boundary string.
The left-invariant Maurer-Cartan one-forms $\w_i(\s)$ are defined by 
\be
\w_i=g^{-1}\del_i g .
\ee
The variation of $g(\s)$ under gauge and modified $\k$-transformations
can be chosen to be
\be
\d_{\BbL} g = g \BbL , \ \ \ \ \ \ \ \ \ \ \D_{\hat{\k}} g = 0, \label{gtransf}
\ee
where $\BbL = \BbL (\BbZ (\s))$.
The coupling of this model to the external gauge fields is 
accomplished by replacing the left-invariant Maurer-Cartan one-form
$\w_i=g^{-1}\del_i g$ by the ``gauged'' version 
\be
g^{-1} D_i \, g = \w_i - \del_i \BbZ^{\BbM} 
\BbA_{\BbM},
\ee
where $D_i$ is the
covariant derivative for the right-action of the gauge group. 

The gauge- and $\k$-invariant action for the open supermembrane is then given 
by \cite{Ceder}
\be
S_{O\!M}  = S_{S\!M}  + S_{W\!Z\!W} , \label{OMaction}
\ee
where $S_{S\!M} $ is the bulk action given in (\ref{SMaction}) and
\bea
S_{W\!Z\!W}  &=& \frac{1}{8 \pi} \int_{\del \Si} d^2 \s \;
\mbox{tr} [\frac{1}{2} \sqrt{-g} g^{ij} ( \w_i - \del_i \BbZ^{\BbM} 
\BbA_{\BbM} ) \cdot ( \w_j - \del_j \BbZ^{\BbN} 
\BbA_{\BbN} )  + \vare^{ij} \del_j \BbZ^{\BbM} 
\w_i \BbA_{\BbM} ] \nn \\
& & - \frac{1}{24 \pi} \int_{\Si} d^3 \hat{\s} \; 
\hat{\vare}^{\hi\hj \hat{k}} \Omega_{\hat{k}\hj\hi} (\hat{\w}) ,
\label{WZWaction}
\eea
where
\be
\Omega_{\hat{k}\hj\hi}(\hat{\w})=
\mbox{tr}(\hat{\w}\wedge\hat{\w}\wedge\hat{\w})_{\hat{k}\hj\hi}.\label{wzw3f}
\ee
The first term in (\ref{WZWaction}) describes the kinetic 
energy for the scalar fields $\phi^a (\s)$ and their interactions with the
pullback of the super-gauge potential $\BbA$ on each of the boundary
strings. The second term is the integral over the membrane of the
Wess-Zumino-Witten three-form, constructed in (\ref{wzw3f}) from a worldvolume
one-form $\hat{\w} = \hat{g}^{-1} d \hat{g}$, where $\hat{g} : \Si \to E_8$.
The map $\hat{g}$ must satisfy
\be
\hat{g} \mid_{\del \Si^{(1)}} = g^{(1)}, \ \ \ \ \ \ \ \ \ \ \ \ \ \ \
\hat{g} \mid_{\del \Si^{(2)}} = g^{(2)}, \label{boundg}
\ee
but is otherwise unspecified. That such a $\hat{g}$ exists will be shown below.
It is straightforward to demonstrate that
the variation of $S_{W\!Z\!W} $ under both gauge 
and local modified $\k$-transformations, $\d_{\BbL}$ and  $\D_{\hat{\k}}$ 
respectively, exactly cancels the variations of the
bulk action $S_{S\!M} $ given in (\ref{SuG}) and (\ref{SuK}) 
provided we choose the parameter $\hat{\k}$ on each boundary to obey
\be
P_- \hat{\k} \equiv \frac{1}{2} (1 - \frac{1}{2\sqrt{-\det g_{ij}}} 
\vare^{ij} \Pi_i^{\, A} \Pi_j^{\, B} \G_{AB} ) \hat{\k} = 0. 
\ee
Note that this is consistent with (\ref{boundBPS}). On the boundary strings 
we can denote $\hat{\k}$ by $\k$. In proving this cancellation, it is
necessary to use the super Yang-Mills constraints on each boundary plane.

We now prove that maps $\hat{g}:\Si\to E_8$ with property (\ref{boundg})
indeed exist. Restoring the boundary index, the two sets of scalar 
fields $\phi^{(n)a}$, one at each boundary, correspond to
two maps $g^{(n)}:\del \Si^{(n)} \to E_8$ with $n=1,2$.
From the \HW point of view, 
these maps are completely independent of each other, as are the $E_8$ gauge
groups that they map to. Now let us assume, as we do later in this paper when 
computing the superpotential, that
\be
\del \Si^{(n)}= \BbC \BbP^1 = S^2
\ee
for $n=1,2$. Since
\be
\pi_2(E_8) = 1,
\ee 
it follows that maps $g^{(1)}$ and $g^{(2)}$ must be homotopically equivalent.
Formally, this means that there exists a continuous map
\be
\hat{g} : S^2 \times I \to E_8 ,
\ee
where $I$ can be taken to be the closed interval $[0,\pi \r ]$, with the 
property that
\be
\hat{g}(\s,0) = g^{(1)} (\s), \ \ \ \ \ \ \ \ \ \ \ \hat{g}(\s,\pi\r) = g^{(2)} 
(\s),
\ee
for $\s \in S^2$. Clearly, however, the membrane manifold is
\be
\Si = S^2 \times I .
\ee
It follows that one can extend the boundary maps to membrane worldvolume maps
\be
\hat{g}: \Si \to E_8 ,
\ee
where
\be
\hat{g} \mid_{\del \Si^{(n)}} = g^{(n)},
\ee
for $n=1,2$, as required.
From this point of view, there appears to be a single $E_8$ gauge group.
Note, however, that one can still vary the boundary maps independently.
This is consistent with the \HW two $E_8$ group interpretation. This result can 
be generalized to $\del \Si^{(n)}$ being any compact Riemann surface.
We conclude that the $\int_{\Si}\Omega (\hat{\w})$ term in (\ref{WZWaction})
can be constructed in a well-defined way.

It will be useful in this paper to discuss some specific properties of this
term. Since $\Omega (\hat{\w})$ is a closed three-form, it 
can be represented locally as $\Omega = d \B$, where $\B (\hat{\w})$ is an 
$E_8$ Lie algebra-valued two-form. If we allow for Dirac-like singularities in 
$\B$, this representation can, with care, be used globally. We can then write
\be
\int_{\Si} \Omega (\hat{\w}) = \int_{\del \Si} \B (\hat{\w}) = \sum_{n=1,2} 
\int_{\del \Si^{(n)}} \B (\w^{(n)}) , \label{wzw}
\ee
where $\B (\w^{(n)})$ is the restriction of $\B$ to the boundary 
$\del \Si^{(n)}$ 
of the membrane and $\w^{(n)} = g^{(n)-1} d g^{(n)}$. Therefore, 
$\int_{\Si} \Omega$, although formally expressed
as an integral over the entire membrane worldvolume, is completely
determined by the values of $\phi^{(n)a} (\s)$ on each of the boundary strings. 
This can also be seen by noticing that small 
variations $ \phi^{(n)a} \to \phi^{(n)a} + \delta \phi^{(n)a}$ which are zero 
on the boundaries leave the action invariant. 
Furthermore, notice that the 
equations of motion derived from the action by using either the left- or the 
right-hand side of (\ref{wzw}) will agree.

\section{BPS Conditions on a Calabi-Yau Threefold:}

We have shown that, in order to preserve
the same supersymmetry as the orbifold fixed planes, a supermembrane
must be oriented along the orbifold direction and terminate on each of the
fixed planes. 
Such an open supermembrane 
must contain chiral Wess-Zumino-Witten fields $\w=g^{-1}dg$ as 
dynamical degrees of freedom. These couple to the boundary background gauge 
fields $\BbA$ in such a way that action (\ref{OMaction}) exhibits 
both $\k$-invariance and gauge invariance. 
Recall that the supermembrane has three unphysical bosonic
degrees of freedom. Hence, one can choose a gauge by specifying three
of the bosonic fields $\hat{X}^{\hat{M}}$ to be explicit functions of the 
supermembrane coordinates $\hat{\s}^{\hi}$ (with $\hi = \hat{0},
\hat{1},\hat{2}$). In the eleven-dimensional \HW backgound, the canonical gauge 
choice was specified in the first line of (\ref{rightconf}). In the curved 
backgrounds of heterotic M-theory, which we will shortly study, 
this gauge can also be chosen.
Here, since the membrane must be oriented in the orbifold direction,
we will take
\be
\hat{X}^{\hat{11}} = \hat{\s}^{\hat{2}},
\ee
leaving the rest of the gauge unspecified for the moment (we will specify the
rest of the gauge in Section 7).

In this paper, we are interested in obtaining an effective four-dimensional
theory with $N=1$ supersymmetry. In particular, we want to compute and study
non-perturbative corrections to the superpotential of such a theory. These
corrections arise from the non-perturbative interaction between the background
and the supermembrane embedded in it. The total action of this theory is
\bea
S_{\mbox{Total}} &=& S_{H\! W} + S_{O\!M}  \nn \\
&=& (S_{S\! G} + S_{Y\! M} ) + ( S_{S\!M}  + S_{W\!Z\!W}  ),
\eea
where $S_{S\! G}$, $S_{Y\! M}$, $S_{S\!M} $ and $S_{W\!Z\!W} $ 
are given in (\ref{11dSG}), (\ref{BoundAction}), (\ref{SMaction}) and 
(\ref{WZWaction}) respectively.

In addition to compactifying on $S^1/\Z_2$, which takes 
eleven-dimensional supergravity to \HW theory, there must also be a 
second dimensional reduction on a real six-dimensional manifold.
This space, which reduces the theory from ten- to four-dimensions on
each orbifold boundary plane, and from eleven- to five-dimensions in the 
bulk space, is taken to be a Calabi-Yau threefold, denoted $CY_3$. 
A Calabi-Yau space is chosen since such a configuration will preserve $N=1$
supersymmetry in four-dimensions. That is, we now consider M-theory and open
supermembranes on the geometrical background
\be
M_{11} = R_4 \times CY_3 \times S^1/\Z_2, \label{background}
\ee
where $R_4$ is four-dimensional flat space.

It is essential that the vacuum state of this theory be Lorentz invariant in 
four-dimensions. Any open supermembrane has an embedding geometry given by
\be
\Si = \CC \times S^1/\Z_2,
\ee
where $\CC$ is a real, two-dimensional surface. Clearly, the requirement of
four-dimensional Lorentz invariance implies that
\be
\CC \subset CY_3 .
\ee
Since $CY_3$ is purely space-like, it follows that we must, henceforth, 
use the Euclidean version of the supermembrane theory.\footnote{Another 
reason to Euclideanize the theory is that,
in this paper, we will perform the calculation of quantum corrections using 
the path-integral formalism.}

In this section, we consider the question of which conditions, if any, are
necessary for the supermembrane theory in such a background to preserve 
an $N=1$ supersymmetry.
The $\k$-transformations (\ref{kappa}) of the superspace coordinates
$ \hbZ^{\hbM} = (\hat{X}^{\hM}, \hat{\T}^{\hat{\mu}})$ do not receive 
boundary corrections. Therefore, equation (\ref{11dBPS}) continues to be 
necessary for preservation of supersymmetry.
For a purely bosonic configuration, this expression becomes
\be
\frac{1}{2}(1-\frac{i}{6\sqrt{\det \hat{g}_{\hi\hj}}}\hat{\vare}^{\hi\hj\hk}
\del_{\hi}\hat{X}^{\hM} \del_{\hj}
\hat{X}^{\hN}\del_{\hk}\hat{X}^{\hP}\hat{\G}_{\hM\hN\hP})\hat{\vare} = 0,
\ee
where $\hat{\vare}$ is a covariantly constant spinor and
we have used (\ref{supermetric}), (\ref{supercoord}) and (\ref{projop}). 
Note that an $i$ appears since we are now in Euclidean space.
The gauge fixing condition
$\hat{X}^{\hat{11}} = \hat{\s}^{\hat{2}}$ reduces this expression to
\be
\frac{1}{2}(1-\frac{i}{2\sqrt{\det g_{ij}}}\vare^{ij}\del_i X^M \del_j X^N 
\G_{MN} \hat{\G}_{\hat{11}}) \hat{\vare} = 0.
\ee
Recall from (\ref{z2chiral}) that, on the boundary planes, the spinor 
$\hat{\vare}$ satisfies $\hat{\G}_{\hat{11}} \hat{\vare} = \hat{\vare}$, 
that is, it has positive ten-dimensional chirality. Therefore, on the boundary
strings, we can denote $\hat{\vare}$ by $\vare$ and write
\be
\frac{1}{2}(1-\frac{i}{2\sqrt{\det g_{ij}}}\vare^{ij}\del_i X^M \del_j X^N 
\G_{MN}) \vare = 0.
\ee

Next, we use the assumption that if $X^M$ describes a coordinate in $R^4$, 
denoted by $y^u$, with $u=6,7,8,9$, then $\del_i y^u =0$. Denote a 
coordinate in $CY_3$ by $\breve{y}^{\breve{U}}$, with 
$\breve{U}=0,1,2,3,4,5$.
Now choose the spinor $\vare$ to be of the form $\vare=\vart \otimes \h $, 
where $\h$ and $\vart$ are 
covariantly constant spinors of $CY_3$ and $R_4$, respectively.
In this case, the above condition becomes
\be
\frac{1}{2}(1-\frac{i}{2\sqrt{\det g_{ij}}}\vare^{ij}\del_i 
\breve{y}^{\breve{U}} \del_j \breve{y}^{\breve{V}} 
\breve{\g}_{\breve{U}\breve{V}} )\h = 0.  \label{supcond}
\ee
where $\breve{\g}_{\breve{U}}=\breve{e}_{\breve{U}}^{\, \breve{K}}
\breve{\g}_{\breve{K}}$, $\breve{e}_{\breve{U}}^{\, \breve{K}}$ are
$CY_3$ sechsbeins and $\breve{\g}_{\breve{K}}$ are the six-dimensional 
Dirac matrices, with flat indices $\breve{K}=0,1,2,3,4,5$.

We now switch notation to exploit the complex structure of the Calabi-Yau space.
The complex coordinates of $CY_3$ are denoted by $\breve{y}^m$ and 
$\breve{y}^{\bm}$
with $m,\bm =1,2,3$. For the K\"{a}hler metric $g_{m\bn}$,  the Clifford
relations for the Dirac matrices take the form
\be
\{\breve{\g}_m , \breve{\g}_n \}=0, \ \ \ \ \  
\{\breve{\g}_{\bm} , \breve{\g}_{\bn} \}=0, 
\ \ \ \ \ \{\breve{\g}_m , \breve{\g}_{\bn} \}=2 g_{m\bn}, \label{cyalgebra}
\ee
It is known \cite{Cand,BeckerBS} that there are two covariant constant spinors 
$\h_-$ and $\h_+$ that can exist on $CY_3$. They can be chosen such that
\be
\breve{\g}_{\bm}\h_+ =0 , \ \ \ \ \ \ \ \ \ \ \breve{\g}_m \h_- = 0 .
\label{etaprop}
\ee
It follows that
\be
\breve{\g}_{m\bn}\h_+ = \frac{1}{2}(\breve{\g}_m\breve{\g}_{\bn} 
-\breve{\g}_{\bn}\breve{\g}_m)\h_+ = -\frac{1}{2}(\breve{\g}_m\breve{\g}_{\bn} 
+\breve{\g}_{\bn}\breve{\g}_m)\h_+ = -g_{m\bn}\h_+ , \label{cysp1}
\ee
and, similarly, that
\be
\breve{\g}_{m\bn}\h_- = g_{m\bn}\h_- . \label{cysp2}
\ee
Here we have chosen a normalization of $\h_{\pm}$ such that $\h_{\pm} = 
\h_{\mp}^{\ast}$. 
In this basis, for an arbitary covariantly constant spinor expressed as
\be
\h = \e^{i\a}\h_+ + \e^{-i\a}\h_- ,
\ee
the condition for unbroken supersymmetry (\ref{supcond}) can be written as
\bea
\h_{\pm} &=& \frac{i}{2\sqrt{\det g_{ij}}}\vare^{ij}(\del_i \breve{y}^m \del_j 
\breve{y}^n \breve{\g}_{mn} + \del_i \breve{y}^{\bm} \del_j 
\breve{y}^n \breve{\g}_{\bm n} \nn \\ 
 & & \ \ \ \ \ \ \ \ \ \ \ \ \ \ \ + \del_i \breve{y}^m \del_j 
\breve{y}^{\bn} \breve{\g}_{m\bn} + \del_i \breve{y}^{\bm} \del_j 
\breve{y}^{\bn} \breve{\g}_{mn}) \h_{\pm} . \label{condforeta}
\eea
Note that there are no terms that mix spinors of different type.

We now write the curve $\CC$ in complex coordinates, defining $z=\s^0+i
\s^1$. The derivatives are
\be
\del_0=\del_z+\del_{\bz}, \ \ \ \ \ \ \ \ \ \ \del_1=i(\del_z-\del_{\bz}).
\ee
Using this and (\ref{etaprop}), conditions (\ref{condforeta}) can be 
rewritten as 
\bea
\h_+ &=& \frac{i}{\sqrt{\det g_{z\bz}}}( 2\del_{\bz} \breve{y}^m \del_z 
\breve{y}^n \breve{\g}_{mn} + 2\del_z \breve{y}^{\bm} \del_{\bz} 
\breve{y}^n \breve{\g}_{\bm n} + \del_{\bz} \breve{y}^{\bm} \del_z 
\breve{y}^n \breve{\g}_{\bm n}) \h_+, \nn \\
\h_- &=& \frac{i}{\sqrt{\det g_{z\bz}}}( 2\del_{\bz} \breve{y}^{\bm} \del_z 
\breve{y}^{\bn} \breve{\g}_{\bm\bn} + 2\del_z \breve{y}^m \del_{\bz} 
\breve{y}^{\bn} \breve{\g}_{m \bn} + \del_{\bz} \breve{y}^m \del_z 
\breve{y}^{\bn} \breve{\g}_{m \bn}) \h_- .
\eea
Since $\breve{\g}_{mn}\h_+$ and $\h_+$ transform differently under 
the holonomy group (and similarly for $\breve{\g}_{\bm\bn}\h_-$ and $\h_-$), 
it follows from (\ref{cysp1}) and (\ref{cysp2}) that
the coefficient of $\breve{\g}_{mn}\h_+$ and $\breve{\g}_{\bm\bn} \h_-$
in each of the above equations has to vanish. That is,
\be
\del_{\bz} \breve{y}^m \del_z \breve{y}^n = 0, \ \ \ \ \ \ \ \ \ \ \ \ \ \ \ 
\del_{\bz} \breve{y}^{\bm} \del_z \breve{y}^{\bn} = 0. \label{BBPPSS}
\ee
These can be satisfied by either a holomorphic curve (for which 
$\del_{\bz} \breve{y}^m = \del_z \breve{y}^{\bn} = 0$)
or an anti-holomorphic curve (for which $\del_z \breve{y}^n =
\del_{\bz} \breve{y}^{\bm} = 0$). 
Suppose the curve is holomorphic. Then we obtain $\h_- = 0$ and no
further conditions on $\h_+$. Therefore, the holomorphic curve leaves the 
supersymmetry generated by $\h_+$ unbroken. Of course, if the curve is
anti-holomorphic, only $\h_-$ survives.
Therefore, the necessary condition for a supermembrane in the background
$M_{11} = R_4 \times CY_3 \times S^1/\Z_2$ to preserve an $N=1$ supersymmetry
when the membrane $\Si = \CC \times S^1/\Z_2$ is embedded as $\CC \subset CY_3$
is that $\CC$ must be either a holomorphic or anti-holomorphic curve.
It is conventional to assume that $\CC$ is holomorphic, thus specifying the 
surviving four-dimensional $N=1$ supersymmetry in terms of the covariantly
constant spinor $\h_+$. We adopt this convention and, in the remainder of this
paper, take $\CC$ to be a holomorphic curve.

\section{Low-Energy Limit and the Heterotic Superstring:}

Thus far, we have shown that for a membrane to be supersymmetric
in the background (\ref{background}), it has to span the interval
$S^1/\Z_2$ and wrap around a holomorphic curve $\CC \subset CY_3$. 
In this section, we take the limit as the radius $\r$ of $S^1$ becomes
small and explicitly compute the small $\r$ limit of the open supermembrane
theory. The result will be the heterotic superstring embedded in ten-dimensional
space
\be
M_{10} = R_4 \times CY_3,
\ee
and wrapped around a holomorphic curve $\CC \subset CY_3$.

We begin by rewriting the action (\ref{OMaction}) for a supermembrane 
with boundary strings as
\bea
S_{O\!M}  &=& T_M \int_{\Si} d^3 \hat{\s} ( 
\sqrt{\det \hat{\Pi}_{\hi}^{\, \hat{A}} \hat{\Pi}_{\hj }^{\, \hat{B}}
\h_{\hat{A}\hat{B}}} - \frac{i}{6} 
\hat{\vare}^{\hi\hj \hat{k}} \hat{\Pi}_{\hi}^{\, \hbA}
\hat{\Pi}_{\hj}^{\, \hbB} \hat{\Pi}_{\hat{k}}^{\, \hbC} 
\hbC_{\hbC\hbB\hbA} ) \nn \\
& & - \frac{1}{8 \pi} \sum_{n=1,2} \int_{\del \Si^{(n)}} 
 d^2 \s \; \{ \mbox{tr}[\frac{1}{2} \sqrt{g}g^{ij} ( \w_i^{(n)} - \BbA^{(n)}_i )
 \cdot ( \w^{(n)}_j - 
\BbA^{(n)}_j ) + i \vare^{ij} \w^{(n)}_i \BbA^{(n)}_j ]\nn \\
& & \ \ \ \ \ \ \ \ \ \ \ \ \ \ \ \ \ \ \ \ \ \ \ \ \
 - i \vare^{ij} \B_{ji} (\w^{(n)}) \} , \label{OMaction2}
\eea
where, again, an $i$ appears multiplying $\vare^{ij}$ as well as
$\hat{\vare}^{\hi\hj \hat{k}}$ because we are in Euclidean space, and we
have used (\ref{wzw}).
Furthermore, it is important to note that the requirement that we
work in Euclidean space changes the sign of each term in (\ref{OMaction2})
relative to the Minkowski signature action (\ref{OMaction}).
The boundary terms describe the gauged chiral Wess-Zumino-Witten model. 
Since they are defined only on the boundary, they are not affected by the
compactification on $S^1/\Z_2$. As for the bulk action,
we identify $\hat{X}^{\hat{11}} = \hat{\s}^{\hat{2}}$ and for all remaining
fields keep only the 
dependence on $\hat{\s}^{\hat{0}}, \hat{\s}^{\hat{1}}$.
The Ansatz for compactification on $S^1/\Z_2$ of the $N=1$ 
eleven-dimensional super-elfbeins is given by
\be
\hbE^{\; \hbA}_{\hbM} = \left( \ba{ccc} 
\hbE^{\; A}_{\BbM} & \hbE_{\BbM}^{\; \hat{11}} & \hbE^{\; \hat{\a}}_{\BbM} \\ 
\hbE^{\; A}_{\hat{11}} & \hbE_{\hat{11}}^{\; \hat{11}} & 
\hbE^{\; \hat{\a}}_{\hat{11}} \ea \right)=
\left( \ba{ccc} \BbE^{\; A}_{\BbM} & \phi \BbV_{\BbM} & 
\BbE_{\BbM}^{\; \hat{\a}}+ \c^{\hat{\a}} \BbV_{\BbM} \\ 0  
& \phi & \c^{\hat{\a}} \ea \right). \label{beinAnsatz}
\ee
Here, $\BbE^{\; \BbA}_{\BbM}=(\BbE^{\; A}_{\BbM},\BbE^{\; \hat{\a}}_{\BbM})$
describes the super-zehnbeins of $N=I\! IA$ ten-dimensional superspace,
$\BbV_{\BbM}$ is a ten-dimensional vector superfield describing a $U(1)$ 
super-gauge field and $\phi$ and $\c^{\hat{\a}}$ are superfields whose leading
components are the dilaton and the dilatino, respectively. Note that we have
made a partial local Lorentz gauge choice by setting $\hbE^{\; A}_{\hat{11}}=0$.
The Ansatz for the super-three-form potential is 
\be
\hbC_{\BbM \BbN \hat{11}} = \BbB_{\BbM \BbN}, \ \ \ \ \ \ \ \ \ \ \ \
\hbC_{\BbM \BbN \BbP} = \BbB_{\BbM \BbN \BbP} . \label{formAnsatz}
\ee

We must also find which conditions are imposed on these quantities by
the $\Z_2$-properties of the eleven-dimensional fields, given in
(\ref{bosuz2}) and (\ref{psiuz2}). We will demand that a superfield have the
same $\Z_2$ transformation properties as its bosonic component.
Note that $\hat{g}_{M\hat{11}}$ in 
(\ref{bosuz2}) has odd parity under $\Z_2$. Since we assume that all
fields are independent of $\hat{X}^{\hat{11}}=\s^{\hat{2}}$,
these components must vanish. Therefore, using (\ref{beinAnsatz}), we obtain
\be
\hat{g}_{M\hat{11}} \equiv \hbE^{\; A}_M \hbE^{\; B}_{\hat{11}} \h_{AB} +
\hbE_M^{\; \hat{11}} \hbE_{\hat{11}}^{\; \hat{11}} = \phi^2 \BbV_M
= 0,
\ee
where we use the same symbol, $\hat{g}_{M\hat{11}}$, for the superfield and
its bosonic metric component.
That is, we must set either $\BbV_{\BbM} =0$ or $\phi=0$. Since $\phi$ is
a diagonal element in the super-elfbeins, $\hbE_{\hat{11}}^{\; \hat{11}}=\phi$,
its vanishing would imply that the determinant of the
induced metric given in (\ref{detindmet}) below is zero. This is unacceptable.
Therefore, we must set 
$\BbV_{\BbM} =0$. The ten-dimensional
three-form $\hbC_{\BbM\BbN\BbP}$ components of $\hbC_{\hbM\hbN\hbP}$ must also
vanish for the same reasons. That is,
\be
\BbB_{\BbA \BbB\BbC} = 0. \label{babczero}
\ee
We are left with the super-zehnbeins $\BbE^{\; \BbA}_{\BbM}=(\BbE^{\; A}_{\BbM},
\BbE^{\; \a}_{\BbM})$, the super-two-form $\BbB_{\BbA\BbB}$, the
dilaton superfield $\phi$ and the dilatino superfield $\c^{\hat{\a}}$.
We note in passing that
\be
\hat{g}_{\hat{11}\hat{11}} = \hbE^{\; A}_{\hat{11}} \hbE^{\; B}_{\hat{11}} 
\h_{AB} + \hbE_{\hat{11}}^{\; \hat{11}} \hbE_{\hat{11}}^{\; \hat{11}} 
= \phi^2 . \label{g1111phi2}
\ee
This relation will be useful in the next section when discussing low energy
moduli fields.

The pullback of the super-elfbeins that enter the first two terms in
(\ref{OMaction2}) become
\bea
\hat{\Pi}_i^{\; A} &=&(\del_i \BbZ^{\BbM})\hbE_{\BbM}^{\; A} +
(\del_i \hat{X}^{\hat{11}} ) \hbE_{\hat{11}}^{\; A} =
(\del_i \BbZ^{\BbM})\BbE_{\BbM}^{\; A} \equiv \Pi^{\; A}_i ,  \nn \\
\hat{\Pi}_i^{\; \hat{11}}&=& (\del_i \BbZ^{\BbM}) \phi \BbV_{\BbM} = 0, \nn \\
\hat{\Pi}_{\hat{11}}^{\; A} &=& (\del_{\hat{11}} \BbZ^{\BbM})
\hbE_{\BbM}^{\; A} = 0, \nn \\
\hat{\Pi}_{\hat{11}}^{\; \hat{11}} &=& \hbE_{\hat{11}}^{\; \hat{11}} = \phi . 
\eea
It is now straightforward to calculate the determinant of the induced
metric. We obtain
\be
\det ( \hat{\Pi}_{\hi}^{\, \hat{A}} \hat{\Pi}_{\hj }^{\, \hat{B}}
\h_{\hat{A}\hat{B}} ) = \det ( \Pi_i^{\, A} \Pi_j^{\, B}
\h_{AB} + \hbE_{\hat{11}}^{\; \hat{11}} \hbE_{\hat{11}}^{\; \hat{11}} )
= \phi^2 \det ( \Pi_i^{\, A} \Pi_j^{\, B} \h_{AB} ). \label{detindmet}
\ee
Finally, we calculate the dimensional reduction of the closed three-form
and find
\be
- \frac{1}{6} 
\hat{\vare}^{\hi\hj \hat{k}} \del_{\hi} \hbZ^{\hbM}
\del_{\hj} \hbZ^{\hbN} \del_{\hat{k}} \hbZ^{\hbP} 
\hbC_{\hbP\hbN\hbM}  =  - \frac{1}{2} \vare^{ij} \del_i \BbZ^{\BbM}
\del_j \BbZ^{\BbN} \BbB_{\BbN\BbM}, \label{CisphiB}
\ee
where we have used (\ref{formAnsatz}) and (\ref{babczero}). Therefore, 
the first part of the action (\ref{OMaction2}) reduces in the small $\r$ limit 
to the string action
\be
S_S  = T_S \int_{\CC} d^2 \s ( \phi 
\sqrt{\det \Pi_i^{\, A} \Pi_j^{\, B}
\h_{AB}} - \frac{i}{2} \vare^{ij} \Pi_i^{\, \BbA}
\Pi_j^{\, \BbB} \BbB_{\BbB\BbA} ),      \label{TypeIAction}
\ee
where 
\be
T_S = T_M \pi \rho \equiv (2 \pi \a')^{-1}
\ee
is the string tension of mass dimension two.

Before we can write the total action for the open supermembrane compactified
on $S^1/\Z_2$, we must discuss the boundary terms in (\ref{OMaction2}).
In the limit that the radius $\r$ of $S^1$ shrinks to zero,
the two orbifold fixed planes coincide. Generically,
the two different boundaries of the supermembrane need not be identified.
However, since our supersymmetric embedding Ansatz assumes all quantities
to be independent of the orbifold coordinate, the two boundary strings
coincide as the zero radius limit is taken.
Putting everything together, we find that the resulting action is
\bea
S_{\CC} &=& T_S \int_{\CC} d^2 \s ( \phi 
\sqrt{\det \Pi_i^{\, A} \Pi_j^{\, B}
\h_{AB}} - \frac{i}{2} \vare^{ij} \Pi_i^{\, \BbA}
\Pi_j^{\, \BbB} \BbB_{\BbB\BbA} ) \nn \\
& & - \frac{1}{8 \pi} \sum_{n=1,2} \int_{\CC} 
 d^2 \s \; \{ \mbox{tr}[\frac{1}{2} \sqrt{g}g^{ij} ( \w_i^{(n)} - \BbA^{(n)}_i )
 \cdot ( \w^{(n)}_j - 
\BbA^{(n)}_j ) + i \vare^{ij} \w^{(n)}_i \BbA^{(n)}_j ]\nn \\
& & \ \ \ \ \ \ \ \ \ \ \ \ \ \ \ \ \ \ \ \ \ \ \ \ \ 
- i \vare^{ij} \B_{ji} (\w^{(n)}) \}, \label{HSaction}
\eea
where
\be
\Pi_i^{\, \BbA} = \del_i \BbZ^{\BbM} \BbE_{\BbM}^{\; \BbA}.
\ee
Note that the gauge supermultiplets $\BbA^{(1)}$ and $\BbA^{(2)}$ now
collectively represent the connection for the single gauge group $E_8 \times
E_8$ pulled back onto $\CC$. We will, henceforth, denote this connection by
$\bar{\BbA}$. Furthermore, $\w^{(1)}$ and $\w^{(2)}$ are now both one-forms
on $\CC$. Define
\be
\bar{g} (\s) = e^{\phi^{(1)a}(\s)T_a^{(1)} +\phi^{(2)a}(\s)T_a^{(2)}},
\label{e8*e8}
\ee
where, collectively, $T_a^{(1)}$ and $T_a^{(2)}$ are the generators of 
$E_8 \times E_8$ and let $\bar{\w} = \bar{g}^{-1} d \bar{g}$. Then, using
the fact that $T_a^{(1)}$ and $T_a^{(2)}$ commute, one can
show that (\ref{HSaction}) can be rewritten as
\bea
S_{\CC} &=& T_S \int_{\CC} d^2 \s ( \phi 
\sqrt{\det \Pi_i^{\, A} \Pi_j^{\, B}
\h_{AB}} - \frac{i}{2} \vare^{ij} \Pi_i^{\, \BbA}
\Pi_j^{\, \BbB} \BbB_{\BbB\BbA} ) \nn \\
& & - \frac{1}{8 \pi} \int_{\CC} 
 d^2 \s \; \{ \mbox{tr}[\frac{1}{2} \sqrt{g}g^{ij} ( \bar{\w}_i - \bar{\BbA}_i )
 \cdot ( \bar{\w}_j - 
\bar{\BbA}_j ) + i \vare^{ij} \bar{\w}_i \bar{\BbA}_j ]\nn \\
& &  \ \ \ \ \ \ \ \ \ \ \ \ \ \ \ \ \ \ \ 
 - i \vare^{ij} \B_{ji} (\bar{\w}) \}. \label{HSaction2}
\eea
Finally, following \cite{Wnab}, we note that if $\CC$ is taken to be the boundary
of some three-ball $\B$, then one can rewrite
\be
\frac{1}{4\pi} \int_{\CC} d^2 \s i \vare^{ij} \B_{ji} (\bar{\w})
= \frac{1}{12\pi} \int_{\B} d^3 \hat{\s} i \hat{\vare}^{\hi\hj\hat{k}}
\Omega_{\hat{k}\hj\hi} (\bar{\w}'), \label{homoext}
\ee
where $\bar{\w}'$ is the homotopic extension of $\bar{\w}$ onto the ball $\B$.
We recognize the action (\ref{HSaction2}) and (\ref{homoext}) 
as that of the heterotic $E_8 \times E_8$ superstring
wrapped on a holomorphic curve $\CC \subset CY_3$.

\section{Superpotential in 4D Effective Field Theory:}

It is essential when constructing the superpotential to have a detailed 
understanding of all the moduli in five-dimensional
heterotic M-theory. Furthermore, we must know
explicitly how they combine to form the moduli of the four-dimensional
low-energy theory. With this in mind, we now briefly review the compactification
of \HW theory to heterotic M-theory on a Calabi-Yau threefold with
$G$-flux. We then further compactify this theory on $S^1/\Z_2$, arriving at
the $N=1$ sypersymmetric action of the effective four-dimensional theory.
We emphasize that, throughout this paper, we take the bosonic components of
all superfields to be of dimension zero, both in five-dimensional heterotic
M-theory and in the associated four-dimensional effective theory.

First consider the compactification from \HW theory to heterotic M-theory.
A peculiar feature of this compactification is that the Bianchi identity for 
the three-form field $C$ is modified due to both gauge and gravitational
anomalies on the boundary fixed planes. This implies that its field-strength 
$G$ has nonzero components in the $CY_3$ direction. As a consequence of this 
nontrivial $G$-flux, the five-dimensional
effective theory of strongly coupled heterotic string theory is given by a
specific gauged version of five-dimensional supergravity. This compactification 
is carried out as follows. Consider the metric
\be
ds_{11}^2=V^{-2 / 3} g_{\hat{u}\hat{v}} d\hat{y}^{\hat{u}} d\hat{y}^{\hat{v}} +
g_{\breve{U}\breve{V}} d \breve{y}^{\breve{U}} d \breve{y}^{\breve{V}},
\label{cymetric}
\ee
where $\hat{y}^{\hat{u}}$, $\hat{u}=6,7,8,9,11$ are the coordinates of the
five-dimensional bulk space of heterotic M-theory, $\breve{y}^{\breve{U}}$,
$\breve{U}=0,1,2,3,4,5$ are the Calabi-Yau coordinates and
$g_{\breve{U}\breve{V}}$ is the metric on the Calabi-Yau space $CY_3$. 
The factor $V^{-2/3}$ in (\ref{cymetric})
has been chosen so that metric $g_{\hat{u}\hat{v}}$ is the five-dimensional
Einstein frame metric.

The non-metric five-dimensional zero-mode fields are obtained by expanding the
fields of eleven-dimensional supergravity in terms of the cohomology classes
of $CY_3$. We state in advance that the five-dimensional hypermultiplets
associated with the $(2,1)$-forms of $CY_3$ cannot contribute to the 
superpotential generated by supermembranes and, hence, will not be discussed
further. Using complex notation, the K\"{a}hler form in $CY_3$ is defined by
\be
\w_{m \bar{n}} = i g_{m \bar{n}}, \label{Kformmetric}
\ee
and can be expanded in terms of the harmonic $(1,1)$-forms $\w_{I m \bar{n}}$,
$I=1,\ldots,h^{1,1}$ as
\be
\w_{m \bar{n}} = \sum_{I=1}^{h^{1,1}} a^I \w_{I m \bar{n}} . \label{Kform}
\ee
The coefficients $a^I = a^I (y^{\hat{u}})$ are the $(1,1)$-moduli of the
Calabi-Yau space. The Calabi-Yau volume modulus $V = V(y^{\hat{u}})$ is
defined by
\be
V = \frac{1}{v} \int_{CY_3} \sqrt{\breve{g}} ,
\ee
where $\breve{g}$ is the determinant of the Calabi-Yau metric 
$g_{\breve{U}\breve{V}}$ and $v$ is a dimensionful parameter necessary to
make $V$ dimensionless. The $h^{1,1}$ moduli $a^I$ and $V$ are not completely
independent. It can be shown that
\be
V = \frac{1}{6} \sum_{I,J,K=1}^{h^{1,1}} d_{IJK} a^I a^J a^K ,
\ee
where coefficients $d_{IJK}$ are the Calabi-Yau intersection numbers defined by
\be
d_{IJK} = \int_{CY_3} \w_I \wedge \w_J \wedge \w_K .
\ee
Therefore, we can take $h^{1,1}-1$ out of the $h^{1,1}$ moduli $a^I$, which we
denote as $a^{\check{I}}$ with $\check{I}=1,\ldots,h^{1,1}-1$, and $V$
as the independent five-dimensional zero-modes.
Now consider the zero-modes of the antisymmetric tensor field 
$\hat{C}_{\hat{M}\hat{N}\hat{P}}$. These are given by 
$\hat{C}_{\hat{u}\hat{v}\hat{w}}$ as well as
\be
\hat{C}_{\hat{u}m\bar{n}} = \sum_{I=1}^{h^{1,1}} \frac{1}{6} A^I_{\hat{u}} 
(y^{\hat{u}}) \w_{Im\bar{n}} \label{CAI}
\ee
and
\be
\hat{C}_{nmp} = \frac{1}{6} \xi (y^{\hat{u}}) \Omega_{mnp},
\ee
where $\Omega_{mnp}$ is the harmonic $(3,0)$-form on $CY_3$.
Therefore, in addition to the graviton $g_{\hat{u}\hat{v}}$, the zero-mode
fields of the five-dimensional effective theory are $h^{1,1} - 1$ real
scalar fields $a^{\check{I}}$, a real scalar $V$, $h^{1,1}$ vector fields 
$A^I_{\hat{u}}$, a complex scalar $\xi$ and $\hat{C}_{\hat{u}\hat{v}\hat{w}}$.

These fields all must be the bosonic components of specific $N=1$ 
supermultiplets in five-dimensions. These supermultiplets are easily identified
as follows.

\noindent 1. Supergravity: the bosonic part of this supermultiplet is
\be
(g_{\hat{u}\hat{v}},\cA_{\hat{u}},\ldots).
\ee
This accounts for $g_{\hat{u}\hat{v}}$ and a linear combination of the vector
moduli $A^I_{\hat{u}}$ which combine to form the graviphoton $\cA_{\hat{u}}$. 
We are left with $h^{1,1} -1$ vector fields, denoted by 
$A_{\hat{u}}^{\check{I}}$.

\noindent 2. Vector supermultiplets: the bosonic part of these supermultiplets 
is
\be
(A^{\check{I}}_{\hat{u}},b^{\check{I}},\ldots).
\ee
Clearly, there are $h^{1,1}-1$ such vector multiplets in the theory, accounting
for the remaining $A^{\check{I}}_{\hat{u}}$ vector moduli. 
The $h^{1,1}-1$ scalars $b^{\check{I}}$ can be identified as
\be
b^{\check{I}} = V^{-1/3} a^{\check{I}} , \label{bVa}
\ee
thus accounting for all $h^{1,1}-1$ $(1,1)$-moduli.

\noindent 3. Universal hypermultiplet: the bosonic part of this supermultiplet 
is
\be
(V, C_{\hat{u}\hat{v}\hat{w}},\xi,\ldots),
\ee
which accounts for the remaining zero-modes discussed above.
Having identified the appropriate $N=1$, five-dimensional superfields, one can
read off the zero-mode fermion spectrum to be precisely those fermions that
complete these supermultiplets.

We now move to the discussion of the compactification of heterotic M-theory in
five-dimensions to the effective $N=1$ supersymmetric theory in four-dimensions.
This compactification was carried out in detail in \cite{B32}. Here we
simply state the resultant four-dimensional zero-modes and their exact
relationship to the five-dimensional moduli of heterotic M-theory.
The bulk space bosonic zero-modes coincide with the $\Z_2$-even fields. One
finds that the metric is
\be
ds_5^2 = R^{-1} g_{uv} dy^u dy^v + R^{2} (d y^{\hat{11}})^2 , \label{5dmetric}
\ee
where $g_{uv}$ is the four-dimensional metric and $R = R(y^u)$ is the
volume modulus of $S^1/\Z_2$. The remaining four-dimensional zero-modes are
\be
V=V(y^u),  \ \ \ \ \ \ \ \ \ \ \ \ b^{\check{I}} = b^{\check{I}} (y^u),
\ee
where ${\check{I}}=1,\ldots,h^{1,1}-1$. In addition, one finds $h^{1,1}$ 
scalar fields
\be
A^I_{\hat{11}} = p^I (y^u) , \label{AIpI}
\ee
$h^{1,1}-1$ of them arising as $A^{\check{I}}_{\hat{11}}$ and one extra
field descending from the eleven-component of the graviphoton $\cA_{\hat{11}}$,
which is $\Z_2$-even. Finally, there is a two-form field
\be
C_{u v \hat{11}} = \frac{1}{3} B_{uv} (y^u).
\ee
This two-form can be dualized to a scalar $\s$ as
\be
H_{uvw}=V^{-2}\e_{\; uvw}^{x} \del_{x} \s ,
\ee
where $H=dB$. It is conventional to redefine these fields into the dilaton
$S$ and $h^{1,1}$ T-moduli $T^I$ as\footnote{Note that the definition of
the imaginary part of $T^I$ differs from that in \cite{B32} by a factor of
$6\sqrt{2}$. The factor chosen here has the same K\"{a}hler potential
as in \cite{B32} and, as we will see, is more natural.}
\be
S = V + i \sqrt{2} \s , \ \ \ \ \ \ \ \ \ \ \ \ \ \ \ 
T^I =  R b^I + i \frac{1}{6} p^I . \label{sandt}
\ee
Note that in the definition of the $T^I$ moduli, we include all $h^{1,1}$
$(1,1)$-moduli, even though they satisfy the constraint
\be
6=\sum_{I,J,K=1}^{h^{1,1}} d_{IJK} b^I b^J b^K .
\ee
This constraint reduces the number of $b^I$ moduli by one, but this is replaced
by the $S^1/\Z_2$ volume modulus $R$. Hence, there remain $2h^{1,1}$ scalar
degrees of freedom from which to form the $h^{1,1}$ $T^I$ chiral 
supermultiplets.
It is then easily seen that these modes form the following four-dimensional,
$N=1$ supermultiplets.

\noindent 1. Supergravity: the full supermultiplet is
\be
(g_{uv},\psi_u^{\a}),
\ee
where $\psi_u^{\a}$ is the gravitino.

\noindent 2. Dilaton and T-moduli chiral supermultiplets: the full multiplets
are
\be
(S,\l_S), \ \ \ \ \ \ \ \ \ \ \ \ \ (T^I, \l_T^I), \label{STmoduli}
\ee
where $I=1,\ldots,h^{1,1}$ and $\l_S$, $\l_T^I$ are the 
dilatino and T-modulinos, respectively.

The fermions completing these supermultiplets arise as zero-modes of the 
fermions of five-dimensional heterotic M-theory. The action for the effective,
four-dimensional, $N=1$ theory has been derived in detail in \cite{B32}. 
Here we simply state the result. The relevant terms for our discussion of
the superpotential are the kinetic terms for the $S$ and $T^I$ 
moduli and the bilinear terms of their superpartner fermions. If we 
collectively denote $S$ and $T^I$ as $Y^{I'}$, where 
$I' = 1,\ldots,h^{1,1}+1$,
and their fermionic superpartners as $\l^{I'}$, then the component
Lagrangian is given by
\bea
\cL_{4D} &=& K_{I'\bar{J}'} \del_u Y^{I'} \del^u 
\bar{Y}^{\bar{J}'} + e^{\k_p^2 K} \left( K^{I'\bar{J}'} 
D_{I'} W \bar{D}_{\bar{J}'} W - 3 \k^2_p | W |^2 \right) \nn \\
& & + K_{I'\bar{J}'} \l^{I'} \delslash \l^{\bar{J}'}
- e^{\k^2_p K / 2} ( D_{I'} D_{J'} W ) \l^{I'} \l^{J'} 
+ \mbox{h.c.}  \label{4Daction}
\eea
Here $\k^2_p$ is the four-dimensional Newton's constant,
\be
K_{I'\bar{J}'} = \del_{I'} \del_{\bar{J}'} K
\ee
are the K\"{a}hler metric and K\"{a}hler potential respectively, and
\be
D_{I'} W = \del_{I'} W + \k^2_p \frac{\del K}{\del Y^{I'}} W
\ee
is the K\"{a}hler covariant derivative acting on the superpotential $W$.
The K\"{a}hler potential was computed in \cite{B32}. In terms of
the $S$ and $T^I$ moduli it is given by
\be
\k^2_p K = - \ln ( S + \bar{S}) - \ln \left( \frac{1}{6} 
\sum_{I,J,K=1}^{h^{1,1}}d_{IJK}(T+\bar{T})^I(T+\bar{T})^J(T+\bar{T})^K \right).
\ee

It is useful at this point to relate the low energy fields of the heterotic
superstring action derived in Section 5 to the four-dimensional moduli
derived here from heterotic M-theory. Specifically, we note from 
(\ref{g1111phi2}) that
\be
\hat{g}_{\hat{11}\hat{11}} \mid_{\T = 0} \; = \, \phi^2 \mid_{\T = 0},
\ee
and from (\ref{cymetric}) and (\ref{5dmetric}) that
\be
d s^2_{11} = \cdots + R^2 V^{-2/3}(d y^{\hat{11}})^2.
\ee
Identifying them implies that
\be
\phi \mid_{\T = 0} \; = R V^{-1/3}\, . \label{phiisR}
\ee
Similarly, it follows from (\ref{CisphiB}), (\ref{CAI}) and (\ref{AIpI})
that
\be
\BbB_{m\bar{n}} \mid_{\T = 0} \; = B_{m\bar{n}} 
= \sum_{I=1}^{h^{1,1}} \frac{1}{6} p^I \w_{Im\bar{n}} . \label{BpI}
\ee
We will use these identifications in the next section.

Following the approach of \cite{BeckerBS} and \cite{HarvMoor}, we will 
calculate the non-perturbative
superpotential by computing instanton induced fermion bilinear interactions 
and then
comparing these to the fermion bilinear terms in the low energy effective 
supergravity
action. In this paper, the instanton contribution arises
from open supermembranes wrapping on
a product of the $S^1/\Z_2$ interval and a holomorphic curve $\CC \subset CY_3$.
Specifically, we will calculate this instanton contribution to the two-point
function of the fermions $\l^I$ associated with the $T^I$
moduli.\footnote{In the remained of this paper, we will drop the subscript
$T$ in $\l_T$ given in (\ref{STmoduli}).}
The two-point function of four-dimensional space-time
fermions $\l^I, \l^J$ located at positions $y^u_1,y^u_2$ is given by 
the following path integral expression
\be
\langle \l^I (y^u_1) \l^J (y^u_2) \rangle  = 
\int \cD  \Phi  e^{-S_{4D}} \l^I (y^u_1) \l^J (y^u_2)
\cdot \int \cD \hbZ \cD \w e^{-S_{\Si}(\hbZ,\w ; \hbE_{\hbM}^{\; \hbA},
\hbC_{\hbM\hbN\hbP},\BbA^{(n)}_{\BbM} )}, \label{pathint}
\ee
where $S_{\Si}$ is the open supermembrane action given in (\ref{OMaction2}).
Here $\Phi$ denotes all supergravity fields in the $N=1$ supersymmetric 
four-dimensional Lagrangian (\ref{4Daction}) and $\hbZ , \w$ are all the 
worldvolume fields on the open supermembrane. In addition,
the path-integral is performed over all supersymmetry preserving 
configurations of the membrane in the eleven-dimensional \HW
background $(\hbE_{\hbM}^{\; \hbA},\hbC_{\hbM\hbN\hbP},\BbA^{(n)}_{\BbM})$ 
compactified down to four-dimensions on $CY_3 \times S^1/\Z_2$. The integration
will restore $N=1$ four-dimensional supersymmetry.
The result of this calculation is then compared to the terms in (\ref{4Daction})
proportional to $(D_I D_J W) \l^I \l^J$
and the non-perturbative contribution to $W$ extracted.

\section{String Action Expansion:}

In this paper, we are interested in the non-perturbative contributions of
open supermembrane instantons to the
two-point function (\ref{pathint}) of chiral fermions in the
four-dimensional effective field theory. In order to preserve $N=1$ 
supersymmetry, the supermembrane must be of the form $\Si = \CC \times 
S^1/\Z_2$, where curve $\CC \subset CY_3$ is holomorphic. As we have shown in 
previous sections, this is equivalent, in the low energy limit,
to considering the non-perturbative contributions of heterotic superstring 
instantons to the same fermion two-point function in the effective 
four-dimensional theory. Of course, in this setting,
the superstring must wrap completely around a holomorphic curve
$\CC \subset CY_3$ in order for the theory to be $N=1$ supersymmetric.

Since we are interested only in non-perturbative corrections to the two-point 
function $\langle \l^I (y^u_1) \l^J (y^u_2) \rangle$, the perturbative 
contributions to this function, which arise from the interaction terms in 
the effective four-dimensional action $S_{4D}$ 
in (\ref{pathint}), will not be considered in this paper. Therefore, we keep
only the kinetic terms of all four-dimensional dynamic fields in $S_{4D}$.
Furthermore, we can perform the functional integrations over all these fields 
except $\l^I$, obtaining some constant determinant factors which we need not
evaluate. Therefore, we can rewrite (\ref{pathint}) as
\bea
\langle \l^I (y^u_1) \l^J (y^u_2) \rangle \; &\propto & \,
\int \cD \l \, e^{-\int d^4 y \sum_{K=1}^{h^{1,1}}
\l^K \delslash  \l^K} \l^I (y^u_1) \l^J (y^u_2) \nn \\
& & \cdot \int \cD {\Bbb{Z}} \cD \w e^{-S_{\CC}(\BbZ,\w ; \BbE_{\BbM}^{\; \BbA},
\BbB_{\BbM\BbN},\phi,\bar{\BbA}_{\BbM} )}, \label{pathint2}
\eea
where $S_{\CC}$ is the heterotic superstring action given in (\ref{HSaction}).
As we will see shortly,
the functional dependence of $S_{\CC}$ on the fields $\l^I$ comes from
the interaction between the superstring fermionic field $\T$
and the ten-dimensional gravitino 
(from which $\l^I$ is derived in the Kaluza-Klein compactification).
Both of these fermions are Weyl spinors in ten-dimensions.\footnote{Note 
that in Euclidean space one does not have Majorana-Weyl spinors in 
ten-dimensions.}

Clearly, to perform the computation of the two-point function
(\ref{pathint2}), we must write the action $S_{\CC}$ in terms of its dynamical 
fields and their interactions with the dimensionally reduced background fields.
This means that we must first expand all superfield expressions in terms of
component fields. We will then expand the action in small fluctuations 
around its extrema 
(solutions to the superstring equations of motion), corresponding
to a saddle-point approximation. We will see that because there exists two 
fermionic zero-modes arising from $\T$, their interaction 
with the gravitino will produce a non-vanishing contribution to
(\ref{pathint2}). Therefore, when performing the path-integrals over
the superstring fields, we must discuss the zero-modes with care.
The next step will be to consider the expression for the
superstring action and to write it in terms of the moduli of the
compactification space $CY_3 \times S^1/\Z_2$.
Finally, we will perform all remaining path integrals in the saddle-point
approximation, obtaining the appropriate determinants. 

This will entail a lengthy calculation. Let us then start by expanding the
ten-dimensional superfields in the action $S_{\CC}$ in terms of the 
component fields.

\subsection*{Expanding in Powers of $\T$:}

We begin by rewriting action $S_{\CC}$ in (\ref{HSaction}) as
\be
S_{\CC} = S_S + S_{W\!Z\!W},  \label{HetStrAction}
\ee
where
\bea
S_S  (\BbZ ; \BbE_{\BbM}^{\; \BbA}(\BbZ),
\BbB_{\BbM\BbN}(\BbZ),\phi(\BbZ)) &=& T_S \int_{\CC} d^2 \s ( \phi 
\sqrt{\det \del_i \BbZ^{\BbM} \BbE_{\BbM}^{\, A}  \del_j \BbZ^{\BbN}
\BbE_{\BbN}^{\, B} \h_{AB}} \nn \\
& & - \frac{i}{2} \vare^{ij} \del_i \BbZ^{\BbM} \BbE_{\BbM}^{\; \BbA}
\del_j \BbZ^{\BbN} \BbE_{\BbN}^{\; \BbB} \BbB_{\BbB\BbA} ) \label{StrAction}
\eea
is the supermembrane bulk action dimensionally reduced on $S^1/\Z_2$ and
\bea
S_{W\!Z\!W} (\BbZ,\w ; \BbA^{(n)}_{\BbM} (\BbZ) ) 
&=& - \frac{1}{8 \pi} \int_{\CC} 
 d^2 \s \; \mbox{tr}[\frac{1}{2} \sqrt{g}g^{ij} ( \bar{\w}_i - \bar{\BbA}_i )
 \cdot ( \bar{\w}_j - 
\bar{\BbA}_j ) + i \vare^{ij} \bar{\w}_i \bar{\BbA}_j ]\nn \\
& & + \frac{1}{24 \pi} \int_{\B} d^3 \hat{\s} i \hat{\vare}^{\hi\hj\hat{k}}
\Omega_{\hat{k}\hj\hi} (\bar{\w}') . \label{wzwAction}
\eea
is the gauged Wess-Zumino-Witten action, where
\be 
\bar{\BbA}_i = \del_i \BbZ^{\BbM} \bar{\BbA}_{\BbM} (\BbZ).
\ee
Note that this action is a functional of $\BbZ (\s)=(X(\s),\T(\s))$.
We now want to expand the superfields in (\ref{HetStrAction}) in powers of
the fermionic coordinate $\T(\s)$. For the purposes of this paper, we need
only keep terms up to second order in $\T$. We begin with $S_S$ given
in (\ref{StrAction}). Using an approach similar to \cite{dewitetal} and
using the results in \cite{Berg}, we find 
that, to the order in $\T$ required, the super-zehnbeins are given by
\be
\BbE_{\BbM}^{\; \BbA} = \left( \ba{cc} E_M^{\, A} - i\bar{\Psi}_M \G^A \T &
\frac{1}{2} \Psi_{M}^{\; \a} + \frac{1}{4}\w_M^{\; CD}(\G_{CD})^{\a}_{\, \n}
\T^{\n} \\ - i \G^A_{\; \m \n} \T^{\n} & \d_{\m}^{\a} \ea \right),
\ee
where $E_M^{\, A} (X(\s))$ are the bosonic zehnbeins, $\Psi(X(\s))$ is the
ten-dimensional gravitino, and $\w_M^{\; CD}(X(\s))$ is the ten-dimensional
spin connection, defined in terms of derivatives of $E_M^{\, A} (X)$.
The super-two-form fields are, up to the required order for the
action to be at most quadratic in $\T$,
\bea
\BbB_{MN} &=& B_{MN} + i \phi ( \bar{\T} \G_{[M} \Psi_{N]} 
+ \frac{i}{4}\bar{\T} \G_{[M} \G^{CD}\T \w_{N]CD} ), \nn \\
\BbB_{M\m} &=& -i \phi (\G_M \T)_{\m} , \nn \\
\BbB_{\m \n} &=& 0,
\eea
where $B_{MN}$ is the ten-dimensional bosonic two-form field. Finally, 
we can write
\be
\phi = R V^{-1/3},
\ee
where we have used (\ref{phiisR}).
Substituting these expressions into action (\ref{StrAction}),
it can be written as
\be
S_S  = S_0 + S_{\T} + S_{\T^2}, \label{StrActionexp}
\ee
where $S_0$ is purely bosonic
\bea
S_0 (X ; E_M^{\, A}(X),B_{MN}(X)) &=& T_S \int_{\CC} d^2 \s ( R V^{-1/3}
\sqrt{\det \del_i X^M \del_j X^N 
E_M^{\, A} E_N^{\, B} \h_{AB}} \nn \\
& & - \frac{i}{2} \vare^{ij} \del_i X^M  \del_j X^N B_{NM} ) , \label{BosAction}
\eea
and $S_{\T}$ and $S_{\T^2}$ are the first two terms (linear and quadratic)
in the $\T$ expansion. Straightforward calculation gives\footnote{In a space
with Minkowski signature, where the spinors are Majorana-Weyl, the fermion 
product would be $\bar{\Psi}_M V^M$. However, in Euclidean space, the
fermions are Weyl spinors only and this product becomes the hermitian sum
$\frac{1}{2}(\bar{\Psi}_M V^M - \bar{V}^M\Psi_M)$.}
\bea
S_{\T} (X, \T ; E_M^{\, A}(X),\Psi_M (X)) &=& 
T_S \int_{\CC} d^2 \s  R V^{-1/3} \sqrt{\det \del_i X^M \del_j X^N 
E_M^{\, A} E_N^{\, B} \h_{AB}} \nn \\
& & \cdot \frac{1}{2}(\bar{\Psi}_M\cV^M-\bar{\cV}^M\Psi_M) \label{vertinAction}
\eea
and
\bea
S_{\T^2} (X,\T; E_M^{\, A}(X)) &=& 
T_S \int_{\CC} d^2 \s  R V^{-1/3} \sqrt{\det \del_i X^M \del_j X^N 
E_M^{\, A} E_N^{\, B} \h_{AB}} \nn \\ & & \cdot(\bar{\T} g^{ij} \G_i D_j \T 
- i \e^{ij} \bar{\T} \G_i D_j \T ) , \label{SFqAction}
\eea
where $D_i \T$ is the covariant derivative 
\be
D_i \T = \del_i \T + \del_i X^N \w^{\ A B}_{N} \G_{A B} \T  , \label{DiT}
\ee
$\G_i$ is the pullback of the eleven-dimensional Dirac matrices
\be
\G_i = \del_i X^M \G_M , \label{Gpullback}
\ee
and $\cV^M$ is the vertex operator for the gravitino $\Psi_M$,
given by
\be
\cV^M = g^{i j} \del_i X^M \del_j X^N \G_N \T 
- i \e^{ij} \del_i X^M \del_j X^N \G_N \T , \label{vertop}
\ee
where $\e^{ij} = \vare^{ij}/\sqrt{g}$.
Now consider the expansion of the superfields in $S_{W\!Z\!W}$ given in 
(\ref{wzwAction}). Here, we need only consider the bosonic part of the expansion
\bea
S_{0W\!Z\!W} (X,\w ; \bar{A}_M (X),E_M^{\, A}(X) ) 
&=& - \frac{1}{8 \pi} \int_{\CC} 
 d^2 \s \; \mbox{tr}[\frac{1}{2} \sqrt{g}g^{ij} ( \bar{\w}_i - \bar{A}_i )
 \cdot ( \bar{\w}_j - 
\bar{A}_j ) + i \vare^{ij} \bar{\w}_i \bar{A}_j ]\nn \\
& & + \frac{1}{24 \pi} \int_{\B} d^3 \hat{\s} i \hat{\vare}^{\hi\hj\hat{k}}
\Omega_{\hat{k}\hj\hi} (\bar{\w}') , \label{boswzw}
\eea
where $\bar{A}_i(\s) = \del_i X^M \bar{A}_M (X(\s))$ is the bosonic pullback of 
$\bar{\BbA}_{\BbM}$. For example, the expansion of $\bar{\BbA}_{\BbM}$ to linear
order in $\T$ contains fermions that are not associated with the moduli of 
interest in this paper. Hence, they can be ignored. Similarly, we can show
that all other terms in the $\T$ expansion of $S_{W\!Z\!W}$ are irrelevant to
the problem at hand.

Note that, in terms of the coordinate fields $X$ and $\T$,
the path integral measure in (\ref{pathint2}) becomes\footnote{Since we are
working in Euclidean space, the spinor fields $\T$ are complex. To be consistent
one must use the integration measure $\cD \bar{\T}\cD \T$. In this paper we 
write the integration measure $\cD \T$ as a shorthand for $\cD \bar{\T}\cD \T$.}
\be
\cD \BbZ \cD \w = \cD X \cD \T \cD \w . \label{measures}
\ee
We can now rewrite the two-point function as
\bea
\langle \l^I (y^u_1) \l^J (y^u_2) \rangle \; &\propto& \,
\int \cD \l \, e^{-\int d^4 y \sum_{K=1}^{h^{1,1}}
\l^K \delslash  \l^K} \l^I (y^u_1) \l^J (y^u_2) \nn \\
& & \cdot \int \cD X \cD \T e^{-(S_0 +S_{\T} +S_{\T^2})} 
\cdot \int \cD \w e^{-S_{0W\!Z\!W}} . \label{pathint2half}
\eea
The last factor
\be
\int \cD \w e^{-S_{0W\!Z\!W}}
\ee
behaves somewhat differently and will be evaluated in the next section.
Here, we simply note that it does not contain the fermion $\l^I$ and, hence,
only contributes an overall determinant to the superpotential. This determinant,
although physically important, does not affect the rest of the calculation,
to which we now turn.
To perform the $X,\T$ path integral, it is essential that we fix
any residual gauge freedom in the $X$ and $\T$ fields.

\subsection*{Fixing the $X$ and $\T$ Gauge:}

First, let us fix the gauge of the bosonic coordinate fields $X$ by 
identifying
\be
X^{m'} (\s ) = \d_i^{m'} \s^i, \label{bosgaug}
\ee
where $m' = 0,1$. This choice, which corresponds to orienting the $X^0$ and
$X^1$ coordinates along the string worldvolume, can always be imposed. This 
leaves eight real bosonic degrees of freedom, which we denote as
\be
X^{m''} (\s ) \equiv y^{m''}(\s ), \label{bosdof}
\ee
where $m''=2,\ldots,9$. Next, let us fix the gauge of the fermionic
coordinate fields $\T$. Recall that $\T$ is a Weyl spinor in ten-dimensional
Euclidean space. Note that there are 
16 complex (or 32 real) independent components in this Weyl spinor. 
Now make an two-eight split in the Dirac matrices
\be
\G_{A} = ( \t_{a'} \otimes \tilde{\g} , 1 \otimes \g_{a''} ) , \label{Gamreduc}
\ee
where $a'=0,1$ and $a''=2,\ldots ,9$ are flat indices, and $\t_{a'}$ and 
$\g_{a''}$ are the two- and eight-dimensional Dirac matrices, 
respectively. Then $\G_{11}\equiv - i\G_0 \G_1 \cdots \G_9$ can be 
decomposed as
\be
\G_{11} = \tilde{\t} \otimes \tilde{\g}
\ee
where $\tilde{\g}=\g_2 \g_3 \cdots \g_9$ and
\be
\tilde{\t} = - i \t_{0} \t_{1} = \left( \begin{array}{cc} 1 & 0 \\
0 & -1 \end{array} \right) . \label{tildetao}
\ee
More explicitly,
\be
\G_{11} = \left( \begin{array}{cc} 
\tilde{\g} & 0 \\ 0 & - \tilde{\g} \end{array} \right) . \label{gamma11}
\ee
In general, the Weyl spinor $\T$ can be written in a generic basis as
\be
\T = \left( \begin{array}{c} \T_1 \\ \T_2 \end{array} \right) . \label{GeneralT}
\ee
Note that $SO(10)$ contains $SO(2)\times SO(8)$ as a maximal subgroup. Under
$SO(8)$, $\T_1$ and $\T_2$ transform independently as spinors.  
The Weyl condition is chosen to be 
\be
\frac{1}{2}( 1 - \G_{11} ) \T = 0 . \label{chir10}
\ee
Using (\ref{gamma11}), this condition implies
\be
\tilde{\g} \T_1 = \T_1 , \ \ \ \ \ \ \ \ \ \ \ \tilde{\g} \T_2 = - \T_2 .
\label{T1T2}
\ee
That is, $\T_1$ ($\T_2$) has positive (negative) eight-dimensional chirality.
It follows from the relation $\G_{11}=\tilde{\t} \otimes \tilde{\g}$ that
the two- and eight-dimensional chiralities of $\T$ are correlated.
Since $\T$ has positive ten-dimensional chirality (\ref{chir10}), this 
implies that the two- and eight-dimensional chiralities are either both 
positive or both negative. That is, $\T$ is in the representation
\be
\mathbf{16}^+ = \mathbf{1}^+ \otimes \mathbf{8}^+ 
\oplus \mathbf{1}^- \otimes \mathbf{8}^-.
\ee
From (\ref{T1T2}), we see that
$\T_1$ is in $\mathbf{1}^+ \otimes \mathbf{8}^+$ and $\T_2$ is in 
$\mathbf{1}^- \otimes \mathbf{8}^-$.

Recall from our discussion of $\k$-symmetry in Section 2 that, because we can
use the $\k$-invariance of the worldvolume theory to gauge away half of the
16 independent components of $\T$, only half of these components
represent physical degrees of freedom. For the superstring, we can define the
projection operators
\be
P_{\pm} = \frac{1}{2} (1 \pm \frac{i}{2\sqrt{g}}\vare^{ij}\Pi_i^{\, A}
\Pi_j^{\, B} \G_{AB})
\ee
and write
\be
\T = P_+ \T + P_- \T .
\ee
Now note from (\ref{kappa}) that $P_+ \T$ can be gauged away, while
the physical degrees of freedom are given by $P_- \T$. Using (\ref{tildetao}), 
it follows that $\T_2$ in (\ref{GeneralT}) can
be gauged to zero, leaving only $\T_1$ as the physical degrees of freedom. 
We thus can fix the fermion gauge so that
\be
\T = \left( \begin{array}{c} \q \\ 0 \end{array} \right), \label{Theta}
\ee
where $\q$ is an $SO(8)$ spinor with positive chirality,
\be
\tilde{\g} \q = \q .
\ee
That is, the physical fermions in the worldsheet theory belong to the 
representation $\mathbf{1}^+ \otimes \mathbf{8}^+$ of $SO(2) \times SO(8)$.

We conclude that the physical degrees of freedom contained in 
$\BbZ = (X,\T)$ are
\be
y^{m''} (\s), \ \ \ \ \ \ \ \ \ \q^{\dot{q}} (\s),
\ee
where $m''=2,\ldots,9$, and $\dot{q}=1,\ldots,8$. The spinor index
$\dot{q}$ corresponds to the positive chirality $SO(8)$-Weyl representation.
Therefore, the $X,\T$ path-integral measure in (\ref{pathint2half}) must be 
rewritten as 
\be
\cD X \cD \T \propto \cD y \cD \q ,
\ee
where there is an unimportant constant of proportionality representing the 
original gauge redundancy.\footnote{Here, again, we write $\cD \q$ as a 
shorthand for $\cD \bar{\q} \cD \q$.}

\subsection*{Equations of Motion:}

We can now rewrite the two-point function (\ref{pathint2half}) as
\bea
\langle \l^I (y^u_1) \l^J (y^u_2) \rangle \; &\propto & 
\int \cD \l \, e^{-\int d^4 y \sum_{K=1}^{h^{1,1}}
\l^K \delslash  \l^K} \l^I (y^u_1) \l^J (y^u_2) \nn \\
& & \cdot \int \cD y \cD \q e^{-(S_0 +S_{\T} +S_{\T^2})} 
\cdot \int \cD \w e^{-S_{0W\!Z\!W}} . \label{pathint3}
\eea
In this paper, we want to use a saddle-point approximation to evaluate these
path-integrals. We will consider small fluctuations $\d y$ and $\d \q$
of the superstring degrees of freedom around a solution $y_0$ and $\q_0$
to the equations of motion
\be
y = y_0 + \d y , \ \ \ \ \ \ \ \ \ \ 
\q = \q_0 + \d \q . \label{allfluc} 
\ee
However, before expanding the action using (\ref{allfluc}), we
need to discuss the equations of motion for the fields $y$ and $\q$, 
as well as their zero-modes.

Consider first the equations of motion for the bosonic fields $y(\s)$. The
bosonic action (\ref{BosAction}) can be written as
\be
S_0 = T_S \int_{\CC} d^2 \s ( R V^{-1/3} \sqrt{\det g_{ij}} 
+ \frac{i}{2} \vare^{ij} b_{ij}  ) , 
\label{BosAct}
\ee
where
\be
g_{ij} = \del_i X^M \del_j X^N g_{MN} , \ \ \ \ \ \ \ \ \ \ \ \ \ \ \ 
b_{ij} = \del_i X^M \del_j X^N B_{MN} . \label{gandb}
\ee
We now assume that the background two-form field $B_{MN}(X)$ satisfies $dB=0$.
This can be done if we neglect corrections of order $\a'$.
Then, locally, $B = d \Lambda$, where $\Lambda$ is a one-form.
Thus the second term in (\ref{BosAct}) can be written as a total derivative
and so does not contribute to the equations of motion. Varying the action,
we obtain the bosonic equations of motion
\bea
\frac{1}{2} \sqrt{\det g_{kl}} g^{ij} \del_i X^M \del_j X^N 
\frac{\del g_{MN}}{\del X^L}
& & \nn \\
- \del_i (\sqrt{\det g_{kl}} g^{ij} \del_j X^M g_{LM}) &=& 0. \label{stringeom}
\eea
where $M,N,L = 0,\ldots ,9$. Since we are considering the product metric on 
$R_4 \times CY_3$, the ten-dimensional metric can be written as
\be
g_{MN} = \left( \ba{cc} g_{\breve{U} \breve{V}} & 0 \\ 0 & \h_{u v} \ea 
\right). \label{gsplit}
\ee
Equation (\ref{stringeom}) then breaks into two parts
\bea
\frac{1}{2} \sqrt{\det g_{kl}} g^{ij} \del_i X^{\breve{U}} 
\del_j X^{\breve{V}} \frac{\del g_{\breve{U}\breve{V}}}{\del X^{\breve{W}}}
& & \nn \\
- \del_i (\sqrt{\det g_{kl}} g^{ij} \del_j X^{\breve{U}} 
g_{\breve{U}\breve{W}}) &=& 0 \label{beom1}
\eea
and
\be
\del_i (\sqrt{\det g_{kl}} g^{ij} \del_j X^u \h_{u v} ) = 0 . \label{beom2}
\ee
It is straightforward to show that the first equation of motion (\ref{beom1})
is consistent with the BPS conditions (\ref{BBPPSS}) obtained in Section 4, 
as they should be. We will consider the second equation shortly.

Next consider the equations of motion for the fermionic degrees of
freedom. In action (\ref{StrAction}) the terms that contain $\T$ are 
(\ref{vertinAction}) and (\ref{SFqAction}), whose sum can be written, taking
into account the gauge fixing condition (\ref{Theta}), as
\be
2 T_S \int_{\CC} d^2 \s R V^{-1/3} \sqrt{\det g_{ij}} 
(\frac{1}{2}(\bar{\Psi}_M V^M - \bar{V}^M\Psi_M) + \bar{\T} \G^i D_i \T ),
\label{sumfact}
\ee
where
\be
V^M = g^{ij} \del_i X^M \del_j X^N \G_N \T .
\ee
It follows from the gauge fixing condition
\be
\T = \left( \begin{array}{c} \q \\ 0 \end{array} \right)
\ee
that only half of the components of $\Psi$ couple to the physical 
degrees of freedom in $\T$, namely 
\be
\Psi^+ = \frac{1}{2} (1 + 1 \otimes \tilde{\t}) \Psi .
\ee
The equations of motion for $\T$ are then found to be
\be
D_{0i} \T_0 = \frac{1}{2} \del_i \breve{y}_0^{\breve{U}} \Psi^+_{\breve{U}}, 
\label{eomforTheta}
\ee
where
\be
D_{0i} \T_0 = \del_i \T_0 + \del_i \breve{y}_0^{\breve{U}} 
\w^{\ \breve{K} \breve{L}}_{\breve{U}} \G_{\breve{K}\breve{L}} \T_0 ,
\ee
and we consider only the physical degrees of freedom $\q_0$ in $\T_0$.

\subsection*{Zero-Modes:}

The saddle-point calculation of the path-integrals
$\cD y$ and $\cD \q$ around a solution to the equations of motion can 
be complicated by the occurrence of zero-modes.
First consider bosonic solutions of the
equations of motion (\ref{stringeom}). By construction, all such solutions
are maps from a holomorphic curve $\CC \subset CY_3$ to the target space
normal to the curve. Since $\CC \subset CY_3$, the four functions which map
to $R_4$, which we denote by $y_0^u$ with $u=6,7,8,9$, are constants
independent of $\s$. Clearly, these can take any value in $R_4$, so we can write
\be
y_0^u \equiv x^u , \label{boszeromodes}
\ee
where $x^u$ are coordinates of $R_4$. Therefore, any solution
of the equations of motion will always have these four translational zero-modes.
These modes are the solution of the second equation of motion (\ref{beom2}).
Are additional zero-modes possible? Generically, the remaining functions
$y_0^U(\s)$, $U=2,3,4,5$ can have other zero-modes. However, to avoid further
technical complications we will, in this paper, consider only curves $\CC$ 
such that
\be
\CC = \BbC \BbP^1 = S^2,
\ee
where the $S^2$ are rigid spheres isolated in $CY_3$. In this case, there are 
clearly no 
additional zero-modes. It follows that for a saddle-point calculation of the
path-integrals around a rigid, isolated sphere the bosonic measure can be
written as
\be
\cD y^{m''} = d^4 x \, \cD \d y^{m''} ,
\ee
where we have expanded
\be
y^{m''} = y^{m''}_0 + \d y^{m''}  \label{bosfluc}
\ee
for small fluctuations $\d y^{m''}$.

Now consider fermionic solutions $\q_0$ of the equation of motion 
(\ref{eomforTheta}). To any $\T_0$ can always be added a solution of the
homogeneous ten-dimensional Dirac equation
\be
D_{0i} \T' = 0 . \label{diraceq}
\ee
This equation has the general solution
\be
\T' = \vart \otimes \h_-  , \label{zeromodeansatz}
\ee
where $\h_-$ is the covariantly constant spinor on $CY_3$, discussed in 
Section 4, which is 
broken by the membrane embedding and $\vart$ is an arbitrary Weyl spinor
satisfying the Weyl equation in $R_4$. Therefore, any solution $\q_0$ of the
equations of motion will always have two complex component fermion
zero-modes $\vart^{\a}$, $\a=1,2$. The rigid, isolated sphere has no additional
fermion zero-modes. Hence, for a saddle-point calculation of the path integrals
around a rigid, isolated sphere the fermionic measure can be written as
\be
\cD \q = d \vart^1 d \vart^2 \, \cD \d \q ,
\ee
where we have expanded
\be
\q = \q_0 + \d \q
\ee
for small fluctuations $\d \q$. To conclude, in the saddle-point approximation
the $y,\q$ part of the path integral measure can be written as
\be
\cD y^{m''}\cD\q =d^4 x\, d \vart^1 d \vart^2 \,\cD\d y^{m''}
\cD\d\q .
\ee

\subsection*{Saddle-Point Calculation:}

We are now ready to calculate the two-point function (\ref{pathint3}), which
can be rewritten as
\bea
\langle \l^I (y^u_1) \l^J (y^u_2) \rangle \; &\propto &
\int \cD \l \, e^{-\int d^4 y \sum_{K=1}^{h^{1,1}}
\l^K \delslash  \l^K} \l^I (y^u_1) \l^J (y^u_2) 
\nn \\
& & \cdot \int d^4 x\, d \vart^1 d \vart^2 \,\cD\d y^{m''}
\cD\d\q \, e^{-(S_0 + S_{\T} + S_{\T^2})} \cdot 
\int \cD \w \,  e^{-S_{0W\!Z\!W}}. \label{pathint4}
\eea
Substituting the fluctuations (\ref{allfluc}) around
the solutions $y_0$ and $\q_0$ into
\be
\S =  S_0 + S_{\T} + S_{\T^2} ,
\ee
we obtain the expansion
\be
\S = \S_0 + \S_2  ,
\ee
where, schematically
\be
\S_0 = \S \mid_{y_0 , \q_0} 
\ee
and
\be
\S_2 = \frac{\d^2 \S}{\d y \d y} \mid_{y_0 , \q_0} (\d y)^2
+ 2 \frac{\d^2 \S}{\d y\d\q} \mid_{y_0 , \q_0} (\d y\d\q)
+ \frac{\d^2 \S}{\d\q\d\q} \mid_{y_0 , \q_0} (\d\q)^2 .
\ee
The terms in the expansion linear in $\d y$ and $\d\q$ each vanish by the
equations of motion. To avoid further complicating our notation, we state in
advance the following simplifying facts. First, note that all terms in $\S_2$
contribute to the two-point function to order $\a'$ on the superstring
worldsheet. Therefore, we should evaluate these terms only to classical order
in $y^{m''}_0$ and $\q_0$. To classical order, one can take $\q_0 = 0$ since,
to this order, the background gravitino on the right-hand side of 
(\ref{eomforTheta}) vanishes. Therefore, $\S_2$ simplifies to
\be
\S_2 = \frac{\d^2 \S}{\d y \d y} \mid_{y_0 , \q_0 = 0} (\d y)^2
+ \frac{\d^2 \S}{\d\q\d\q} \mid_{y_0 , \q_0 = 0} (\d\q)^2 .
\ee
It is useful to further denote
\be
\S_0 = \S_0^y + \S_0^{\q} , 
\ee
where
\be
\S_0^y = ( S_0 ) \mid_{y_0}, \ \ \ \ \ \ \ \ \ \ \ \ \
\S_0^{\q} = ( S_{\T} + S_{\T^2} ) \mid_{y_0,\q_0} , \label{sysq}
\ee
and to write
\be
\S_2 = \S_2^y + \S_2^{\q},
\ee
with
\be
\S_2^y = \frac{\d^2 \S}{\d y \d y} \mid_{y_0 , \q_0 =0} (\d y)^2 ,
\ \ \ \ \ \ \ \ \ \ \
\S_2^{\q} = \frac{\d^2 \S}{\d\q\d\q} \mid_{y_0 , \q_0 =0} (\d\q)^2 . \label{s2}
\ee
We can then rewrite two-point function (\ref{pathint4}) as
\bea
\langle \l^I (y^u_1) \l^J (y^u_2) \rangle \; &\propto &
\int \cD \l \, e^{-\int d^4 y \sum_{K=1}^{h^{1,1}}
\l^K \delslash  \l^K} \l^I (y^u_1) \l^J (y^u_2) \nn \\
& & \cdot \int d^4 x\, e^{-\S_0^y} \cdot \int d \vart^1 d \vart^2 \, 
e^{-\S_0^{\q}} \nn \\ 
& & \cdot \int \cD\d y^{m''} \, e^{-\S_2^y} \cdot
\int \cD\d\q \, e^{-\S_2^{\q}} \cdot 
\int \cD \w \,  e^{-S_{0W\!Z\!W}}. \label{pathint5}
\eea
We will now evaluate each of the path-integral factors in this expression one
by one. We begin with $\int d^4 x \, e^{-\S_0^y}$.

\subsection*{The $\S_0^y$ Term:}

It follows from (\ref{sysq}) that $\S_0^y$ is simply $S_0$, given in
(\ref{BosAct}) and (\ref{gandb}), evaluated at a solution of the
equations of motion $y_0^{m''}$. Using (\ref{boszeromodes}), which 
implies that $\del_i y_0^u = 0$ for coordinates $y_0^u$ of $R_4$, and the
form of the ten-dimensional metric
\be
ds_{10}^2 =  g_{uv} dy^u dy^v + g_{m \bar{n}} d\breve{y}^m 
d\breve{y}^{\bar{n}},
\ee
with $\breve{y}^m,\breve{y}^{\bar{m}}$ complex coordinates of $CY_3$, we see
that
\be
\S_0^y = T_S \int_{\CC} d^2 \s ( R V^{-1/3} \sqrt{\det g_{ij}} + \frac{i}{2} 
\vare^{ij} b_{ij} ), \label{threechecks}
\ee
where
\be
g_{ij} = \del_i \breve{y}_0^m \del_j \breve{y}_0^{\bar{n}} g_{m\bar{n}} ,
\ \ \ \ \ \ \ \ \ \ \ \ \
b_{ij} = \del_i \breve{y}_0^m \del_j \breve{y}_0^{\bar{n}} B_{m\bar{n}} .
\label{gandb2}
\ee
Let us evaluate the term involving $g_{ij}$. To begin, we note that
\be
\int_{\CC} d^2 \s \sqrt{\det g_{ij}} = \frac{1}{2} \int_{\CC} d^2 \s \sqrt{g}
g^{ij} \del_i \breve{y}_0^m \del_j \breve{y}_0^{\bar{n}} g_{m\bar{n}} , 
\label{onecheck}
\ee
where the first term is obtained from the second using the worldvolume metric
equation of motion. Noting that $g_{ij}$ is conformally flat, and going to
complex coordinates $z=\s^0 + i \s^1$, $\bar{z}=\s^0 - i\s^1$, it follows
from (\ref{onecheck}) that
\be
\int_{\CC} d^2 \s \sqrt{\det g_{ij}} = \frac{1}{2} \int_{\CC} d^2 z
\del_z \breve{y}_0^m \del_{\bar{z}} \breve{y}_0^{\bar{n}} \w_{m\bar{n}} , 
\label{twochecks}
\ee
where $\w_{m\bar{n}}=ig_{m\bar{n}}$ is the K\"{a}hler form on $CY_3$. In 
deriving (\ref{twochecks}) we used the fact, discussed in Section 4, that
the functions $\breve{y}^m$ must be holomorphic. Recall from (\ref{Kform})
that
\be
\w_{m \bar{n}} = \sum_{I=1}^{h^{1,1}} a^I \w_{I m \bar{n}} .
\ee
Therefore, we can write
\be
R V^{-1/3} \int_{\CC} d^2 \s \sqrt{\det g_{ij}} = \frac{v_{\CC}}{2} 
\sum_{I=1}^{h^{1,1}} R b^I \w_I ,
\ee
where $b^I = V^{-1/3} a^I$,
\be
\w_I = \frac{1}{v_{\CC}} \int_{\CC} d^2 z \, \w_{Iz\bar{z}} \label{omegaI}
\ee
and
\be
\w_{Iz\bar{z}} = \del_z \breve{y}^m \del_{\bar{z}} 
\breve{y}^{\bar{n}} \w_{Im\bar{n}} \label{wIzz}
\ee
is the pullback onto the holomorphic curve $\CC$ of the $I$-th harmonic
$(1,1)$-form. Note that we have introduced the parameter $v_{\CC}$ of mass
dimension minus two to make $\w_I$ dimensionless. Parameter $v_{\CC}$ can
naturally be taken to be the volume of curve $\CC$.
Now consider the second term in (\ref{threechecks})
involving $b_{ij}$. First, we note that
\be
\int_{\CC} d^2 \s  \frac{i}{2} \vare^{ij} b_{ij} = \frac{i}{2} \int_{\CC} d^2 z
\del_z \breve{y}_0^m \del_{\bar{z}} \breve{y}_0^{\bar{n}} B_{m\bar{n}} .
\ee
Remembering from (\ref{BpI}) that
\be
B_{m\bar{n}} = \sum_{I=1}^{h^{1,1}} \frac{1}{6} p^I \w_{Im\bar{n}},
\ee
it follows that
\be
\int_{\CC} d^2 \s  \frac{i}{2} \vare^{ij} b_{ij} = \frac{v_{\CC}}{2}
\sum_{I=1}^{h^{1,1}} \frac{i}{6} p^I \w_I .
\ee
Putting everything together, we see that
\be
\S_0^y = \frac{T}{2} \sum_{I=1}^{h^{1,1}} \w_I (R b^I + \frac{i}{6} p^I ),
\ee
where
\be
T = T_S v_{\CC} = T_M \pi \rho \, v_{\CC}
\ee
is a dimensionless parameter.
Recalling from (\ref{sandt}) the $T^I$ moduli are defined by
\be
T^I = R b^I + i \frac{1}{6} p^I ,
\ee
it follows that we can write $\S_0^y$ as
\be
\S_0^y = \frac{T}{2} \sum_{I=1}^{h^{1,1}} \w_I T^I .
\ee
We conclude that the $\int d^4 x e^{-\S_0^y}$ factor in the path-integral
is given by
\be
\int d^4 x \, e^{-\S_0^y} = \int d^4 x e^{-\frac{T}{2}\sum_{I=1}^{h^{1,1}}
\w_I T^I } .
\ee
We next evaluate the path integral factor $\int d \vart^1 d \vart^2 
e^{-\S_0^{\q}}$.

\subsection*{The $\S_0^{\q}$ Term and the Fermionic Zero-Mode Integral:}

It follows from (\ref{sysq}) that $\S_0^{\q}$ is the sum of $S_{\T}$ and
$\S_{\T^2}$, given in (\ref{sumfact}), evaluated at a solution of the
equations of motion $y_0^{m''}$, $\q_0$. Varying (\ref{sumfact}) with
respect to $\bar{\T}$ leads to the equation of motion $D_i \T_0 = \frac{1}{2}
\del_i X_0^M \Psi_M$, where we have used (\ref{vertop}) and the fact that
the Dirac matrices can be taken to be hermitean. Inserting the equation of 
motion into (\ref{sumfact}) we find
\be
\S_0^{\q} = T_S \int_{\CC} d^2 \s R V^{-1/3} \sqrt{\det g_{ij}}
\bar{\Psi}_M V_0^M ,
\ee
where 
\be
V_0^M = g^{ij} \del_i X_0^M \del_j X_0^N \G_N \T_0 .
\ee
Recalling that $\del_i y_0^u = 0$ for all coordinates $y_0^u$ of $R_4$,
it follows that the only non-vanishing components of $V_0^M$ are
\be
V_0^{\breve{U}} = g^{i j} \del_i \breve{y}_0^{\breve{U}} \del_j 
\breve{y}_0^{\breve{W}} \G_{\breve{W}} \T_0 ,
\ee
where $g_{ij}$ is given by (\ref{gandb2}). We conclude that
\be
\S_0^{\q} = T_S \int_{\CC} d^2 \s R V^{-1/3} \sqrt{\det g_{ij}}
\bar{\Psi}_{\breve{U}} V_0^{\breve{U}} . \label{cyvertop}
\ee
Now $V_0^{\breve{U}} \propto \T_0$, where $\T_0$ satisfies the equation of
motion (\ref{eomforTheta}). As discussed above, any such solution can be
written as the sum
\be
\T_0 = \hat{\T}_0 + \T' ,
\ee
where $\T'$ is a solution of the purely homogeneous Dirac equation 
(\ref{diraceq}) and has the form (\ref{zeromodeansatz}). Since in the
path-integral we must integrate over the two zero-modes $\vart^{\a}$,
$\a=1,2$ in $\T'$, it follows that terms involving $\hat{\T}_0$ can never
contribute to the fermion two-point function. Therefore, when computing
the superpotential, one can simply drop $\hat{\T}_0$. Hence, $\S_0^{\q}$ is
given by (\ref{cyvertop}) where
\be
V_0^{\breve{U}} = g^{i j} \del_i \breve{y}_0^{\breve{U}} \del_j 
\breve{y}_0^{\breve{W}} \G_{\breve{W}} \T' .
\ee
Inserting expression (\ref{zeromodeansatz}), and using the decomposition
$\G_{\breve{W}} = 1 \otimes \breve{\g}_{\breve{W}}$, implies that
\be
V_0^{\breve{U}} = g^{i j} \del_i \breve{y}_0^{\breve{U}} 
\del_j \breve{y}_0^{\breve{W}} 
\vart\otimes(\breve{\g}_{\breve{W}} \h_-) . \label{cyvert}
\ee
Next, we note that the Kaluza-Klein Ansatz for the ten-dimensional gravitino
in the Calabi-Yau directions is given by
\be
R V^{-1/3}\Psi_{\breve{U}} = - \sum_{L=1}^{h^{1,1}} i 
\omega_{L\breve{U}\breve{V}}  
\l^L \otimes (\breve{\g}^{\breve{V}}\h_+), \label{cygravit}
\ee
where $\omega_{L\breve{U}\breve{V}} (\breve{y}^{\breve{U}})$, 
$L=1,\ldots,h^{1,1}$ are the harmonic $(1,1)$-forms on $CY_3$,
$\l^L(y^u)$ are the fermionic superpartners of the moduli $T^L$
and $\h_+(\breve{y}^{\breve{U}})$ is the Calabi-Yau covariantly constant spinor.
Note that the left-hand side of (\ref{cygravit}) includes a factor of
$R V^{-1/3}$, while the right-hand side has a factor of $-i$ which makes 
the Ansatz consistent with the compactification
moduli defined in (\ref{Kformmetric}), (\ref{bVa}) and (\ref{sandt}).
Using (\ref{cyvert}) and (\ref{cygravit}), one can evaluate the product
$\bar{\Psi}_{\breve{U}} V_0^{\breve{U}}$, which is found to be
\be
R V^{-1/3} \bar{\Psi}_{\breve{U}} V_0^{\breve{U}} = -i g^{i j} 
\del_i \breve{y}_0^{\breve{U}} \del_j \breve{y}_0^{\breve{W}} 
\sum_{L=1}^{h^{1,1}}
\omega_{L\breve{U}\breve{V}} (\h_+^{\dagger}\breve{\g}^{\breve{V}}
\breve{\g}_{\breve{W}} \h_-) \cdot (\l^L \vart) .
\ee
where $\l^L \vart = \l^L_{\a} \vart^{\a}$. Substituting this expression into
(\ref{cyvertop}) then gives
\be
\S_0^{\q} = T \sum_{L=1}^{h^{1,1}} \w_L \, \l^L \vart
\ee
where we have used complex coordinates $z=\s^0 + i \s^1$, $\bar{z}=\s^0 -i\s^1$
for the holomorphic curve, as well as $\breve{y}^m,\breve{y}^{\bar{m}}$ for the
Calabi-Yau coordinates $\breve{y}^{\breve{U}}$. We have also used the property
\be 
\h_+^{\dagger} \breve{\g}^{\bm} \breve{\g}_{\bn} \h_- = 2 \d^m_n
\ee
of the covariantly constant spinors on $CY_3$, derived from 
(\ref{cyalgebra})--(\ref{cysp1}). The coefficients $\w_L$ are given in 
(\ref{omegaI}).
It follows that the $\int d \vart^1 d \vart^2 \, e^{-\S_0^{\q}}$ factor in
the path-integral is
\be
\int d \vart^1 d \vart^2 e^{- \S_0^{\q}} = \int d \vart^1 d \vart^2 
e^{ - T \sum_{L=1}^{h^{1,1}} \w_L \, \l^L \vart } .
\ee
Expanding the exponential, and using the properties of the Berezin integrals, we
find that
\be
\int d \vart^1 d \vart^2 e^{- \S_0^{\q}} = \frac{T^2}{2} \sum_{L,M=1}^{h^{1,1}}
\w_L \w_M \, \l^L \l^M ,
\ee
where we have suppressed the spinor indices on $\l^L \l^M$.

Collecting the results we have obtained thus far, two-point function 
(\ref{pathint5}) can now be written as
\bea
\langle \l^I (y^u_1) \l^J (y^u_2) \rangle \; &\propto &
\int \cD \l \, e^{-\int d^4 y \sum_{K=1}^{h^{1,1}}
\l^K \delslash  \l^K} \l^I (y^u_1) \l^J (y^u_2) \nn \\ 
& & \cdot \int d^4 x\, e^{- \frac{T}{2}\sum_{I=1}^{h^{1,1}}\w_I T^I (x)}
\sum_{L,M=1}^{h^{1,1}} \w_L \w_M \, \l^L (x) \l^M (x) \nn \\ 
& & \cdot \int \cD\d y^{m''} \, e^{-\S_2^y} \cdot
\int \cD\d\q \, e^{-\S_2^{\q}} \cdot 
\int \cD \w \,  e^{-S_{0W\!Z\!W}}. \label{pathint6}
\eea
Next, we evaluate the bosonic path-integral factor $\int \cD \d y^{m''}
e^{- \S_2^y}$.

\subsection*{The $\S_2^y$ Quadratic Term:}

It follows from (\ref{s2}) that $\S_2^y$ is simply the quadratic term in
the $y=y_0 + \d y$ expansion of $\S_0$, given in (\ref{BosAct}) and 
(\ref{gandb}). Note that $S_{\T} + S_{\T^2}$ does not contribute since the 
second derivative is to be evaluated for $\q_0 = 0$. Furthermore, since this
contribution to the path-integral is already at order $\a'$, $\S_0$ should
be evaluated to lowest order in $\a'$. As discussed above, to lowest order
$dB=0$ and, hence, the $b_{ij}$ term in (\ref{BosAct}) is a total divergence
which can be ignored. Within thie bosonic gauge (\ref{bosgaug}), 
choose a system of coordinates such that the metric tensor restricted to
the holomorphic curve $ \CC $ can be written locally as
\be
g_{MN} \mid_{\CC} = \left( \ba{cc} h_{m' n'} (\s) & 0 \\ 0 & h_{m'' n''} (\s)
\ea \right) .
\ee
Performing the expansion, we find that
\be
S_2^y = T_S \int_{\CC} d^2 \s R V^{-1/3} \sqrt{\det g_{ij}} \left( \frac{1}{2} 
g^{ij} (D_i \d y^{m''} )(D_j \d y^{n''})h_{m'' n''}
- \d y^{m''} {\cal U}_{m'' n''} \d y^{n''} \right), \label{s2y}
\ee
where the induced covariant derivative of $\d y$ is given by
\be
D_i \d y^{m''} = \del_i \d y^{m''} + \w_{i \; \, n''}^{m''} \d y^{n''}
\label{indcov}
\ee
and the mass matrix is
\be
{\cal U}_{m'' n''} = \frac{1}{2} R^{m'}_{\; \, m'' m' n''} + 
\frac{1}{8}Q^{m' n'}_{\; \; \; \, m''}Q_{m' n' n''},
\ee
where $R^{m'}_{\; \, m'' m' n''}$ is the ambient curvature tensor restricted 
to the string and $Q_{m' n' n''}$ is the second fundamental 
form.\footnote{These are standard results. 
For a proof, see for example \cite{chernlec}.} 
Note that the eight quantities $\d y^{m''}$ are scalars from the point of
view of the curve $\CC$. Four of these scalars form a 
vector in the tangent bundle of $CY_3$ and the remaining four form a vector 
in $R_4$.
By definition, metric $h_{m'' n''}$ can be decomposed as
\be
h_{m'' n''} = \left( \ba{cc} \h_{uv} & 0 \\ 0 & h_{UV} (\s) 
\ea 
\right) ,
\ee
where $\h_{uv}$ is the flat metric of $R_4$ and $h_{UV}$ is the metric of the
normal space $CY_{\perp}$ defined by the four directions of $CY_3$ that are
perpendicular to the curve $\CC \subset CY_3$. It follows that all connection
components $\w_{i\; n''}^{\, m''}$ with either $m''$ and/or $n''$ in $R_4$ will
vanish. Then (\ref{s2y}) becomes
\bea
S_2^y &=& T_S \int_{\CC} d^2 \s  R V^{-1/3}
\sqrt{\det g_{ij}} \, \{ \frac{1}{2} g^{ij} (\del_i \d y^u)(\del_j \d y^v)
\h_{uv} \nn \\ & & + \frac{1}{2} g^{ij} (D_i \d y^U )(D_j \d y^V) h_{UV}
- \d y^U {\cal U}_{UV} \d y^V \}.
\eea
Integrating the derivatives by parts then gives
\bea
S_2^y &=& T_S \int_{\CC} d^2 \s  R V^{-1/3}
\{ - \frac{1}{2} \d y^u [ \h_{uv} \sqrt{g} g^{ij} \cD_i \del_j ] \d y^v
 \nn \\ & & - \frac{1}{2} \d y^U [ \sqrt{g} ( g^{ij} \cD_i h_{UV} D_j 
+ 2 \, {\cal U}_{UV} ) ] \d y^V \} ,
\eea
where the symbol $\cD_i$ indicates the covariant derivative with respect to both
the induced connection in (\ref{indcov}) and the worldvolume connection of
$\CC$. Generically, the fields $R,V$ are functions of $x^u$. However,
as discussed above, at the level of the quadratic contributions to the
path-integrals all terms should be evaluated at the classical values of
the background fields. Since $R$ and $V$ are moduli, these classical values
can be taken to be constants, rendering $RV^{-1/3}$ independent of $x^u$.
Hence, the factor $T_S RV^{-1/3}$ can simply be absorbed by a redefinition of
the $\d y$'s. Using the relation
\be
\int \cD \d y \; e^{-\frac{1}{2} \int d^2 \s \, \d y \cO \d y} \propto
\frac{1}{\sqrt{\det \cO}} ,
\ee
we conclude that
\be
\int \cD d y^{m''} e^{- \S_2^y} \propto \frac{1}{\sqrt{\det \cO_1}} \,
\frac{1}{\sqrt{\det \cO_2}} 
\ee
where
\bea
\cO_1 &=& \h_{uv} \sqrt{g} g^{ij} \cD_i \del_j , \nn \\
\cO_2 &=& \sqrt{g} (g^{ij} \cD_i h_{UV} D_j + 2 \, {\cal U}_{UV}). \label{Os}
\eea
We next turn to the evaluation of the $\int \cD \d \q \,e^{- \S_2^{\q}}$ factor
in the path-integral.

\subsection*{The $\S_2^{\q}$ Quadratic Term:}

It follows from (\ref{s2}) that $\S_2^{\q}$ is the quadratic term in the
$\q = \q_0 + \d \q$ expansion of $S_{\T^2}$, given in (\ref{SFqAction}).
Note that $S_0 + S_{\T}$ does not contribute. Performing the expansion and
taking into account the gauge fixing condition (\ref{Theta}), we find that
\be
\S_2^{\q} = 2T_S \int_{\CC} d^2 \s R V^{-1/3} \sqrt{\det g_{ij}}
\d \bar{\T} \G^i D_i \d \T . \label{quadtheta}
\ee
One must now evaluate the product $\d\bar{\T} \G^i D_i \d\T$ in terms of the
gauged-fixed quantities $\d\q$. We start by rewriting
\bea
\d \bar{\T} \G^i D_i \d \T &=& g^{ij} \del_j X^M \d \bar{\T} \G_M \del_i \d \T 
\nn \\
& & + g^{ij} \del_j X^M \del_i X^N \w^{\ A B}_{N} 
\d \bar{\T} \G_M \G_{A B} \d \T , \label{Thet21}
\eea
where we have used (\ref{DiT}) and (\ref{Gpullback}).
After fixing the gauge freedom of the bosonic fields $X^M (\s)$
as in (\ref{bosgaug}), expression (\ref{Thet21}) becomes
\bea
\d \bar{\T} \G^i D_i \d \T &=& g^{ij} \d_j^{m'} e_{m'}^{\ a'} \d \bar{\T} 
\G_{a'} \del_i \d \T + g^{ij} e_{m''}^{\ a''} \del_j y^{m''} \d \bar{\T} 
\G_{a''} \del_i \d \T \nn \\
& & + g^{ij} \d_j^{m'} e_{m'}^{\ a'} ( \d_i^{n'} \w^{\ A B}_{n'} + \del_i 
y^{m''} \w^{\ A B}_{m''} ) \d \bar{\T} \G_{a'} \G_{A B} \d \T \nn \\
& & + g^{ij} e_{m''}^{\ a''} \del_j y^{m''} ( \d_i^{m'} \w^{\ A B}_{m'} + 
\del_i y^{n''} \w^{\ A B}_{n''} ) \d \bar{\T} \G_{a''} \G_{A B} \d \T ,
\label{Thet22}
\eea
where $A=(a',a'')$. We see that we must evaluate the fermionic products
\be
\d \bar{\T} \G_{a'} \del_i \d \T, \ \ \ \ \d \bar{\T} \G_{a''} \del_i \d \T, 
\ \ \ \ \d \bar{\T} \G_{a'} \G_{A B} \d \T, \ \ \ \ \d \bar{\T} \G_{a''} 
\G_{A B} \d \T 
\ee
in terms of $\d \q$.
After fixing the fermionic gauge according to (\ref{Theta}),
we can compute the relevant terms in the expression (\ref{Thet22}). 
Consider a product of the type 
$\d \bar{\T}M\d\T$, where $M$ is a $32\times 32$ matrix-operator,
\be
M = \left( \ba{cc} M_1 & M_2 \\ M_3 & M_4 \ea \right) .
\ee
Using (\ref{Gamreduc}) and (\ref{Theta}), we have
\be
\d \bar{\T} M \d \T = \d \T^{\dagger} M \d \T =  \d \q^{\dagger} M_1 \d \q
\ee
Therefore, using (\ref{Gamreduc}), we have the following results
\be
\ba{rl} \d \bar{\T}\G_{a'}\del_i\d\T= 0,
 & \d \bar{\T}\G_{a''}\del_i\d\T=\d\q^{\dagger}\g_{a''}\del_i\d\q , \\
\d\bar{\T}\G_{a'}\G_{b' c'}\d\T=0, & 
\d \bar{\T}\G_{a'}\G_{b' a''}\d\T=(\d_{a' b'}-i\vare_{a' b'})\d\q^{\dagger}
\g_{a''}\d\q , \\
\d\bar{\T}\G_{a'}\G_{a'' b''}\d\T=0, &
 \bar{\T}\G_{a''}\G_{a' b'}\T=-i\vare_{a' b'}\d\q^{\dagger}\g_{a''}\d\q , \\
\d\bar{\T}\G_{a''}\G_{a' b''}\d\T=0, & 
\d \bar{\T}\G_{a''}\G_{b'' c''}\d\T=\d\q^{\dagger}\g_{a''}\g_{b'' c''}\d\q .\ea
\ee
Substituting these expressions into (\ref{Thet22}) yields
\bea
\d\bar{\T}\G^i D_i\d\T &=&  
g^{ij} e_{m''}^{\ a''} \del_j y^{m''} [ \d\q^{\dagger}\g_{a''}\del_i\d\q
-i \w_i^{\ a' b'} \vare_{a' b'} \d\q^{\dagger}\g_{a''}\d\q
+ \w_i^{\ b'' c''} \d\q^{\dagger}\g_{a''}\g_{b'' c''}\d\q ] \nn \\
& & + g^{ij}\d_j^{m'}e_{m'}^{\ a'} \w_i^{\ b' a''} 
(\d_{a' b'}-i\vare_{a' b'})\d\q^{\dagger} \g_{a''}\d\q 
\eea
where
\bea
\w^{\ a' b'}_i &=& \d_i^{m'}\w^{\ a' b'}_{m'}+\del_i y^{n''}
\w^{\ a' b'}_{n''} , \nn \\
\w^{\ b'' c''}_i &=& \d_i^{m'}\w^{\ b'' c''}_{m'}+\del_i y^{n''}
\w^{\ b'' c''}_{n''} , \nn \\
\w^{\ b' a''}_i &=& \d_i^{m'}\w^{\ b' a''}_{m'}+\del_i y^{n''}
\w^{\ b' a''}_{n''} .
\eea
Then (\ref{quadtheta}) becomes
\bea
\S_2^{\q} &=& 2T_S \int_{\CC} d^2 \s R V^{-1/3} 
\d\q^{\dagger} \{ \sqrt{g} g^{ij} e_{m''}^{\ a''} \del_j y^{m''} 
[ \g_{a''}\del_i -i \w_i^{\ a' b'} \vare_{a' b'} \g_{a''} \nn \\
& & + \w_i^{\ b'' c''} \g_{a''}\g_{b'' c''}]
+ \sqrt{g} g^{ij} \d_j^{m'}e_{m'}^{\ a'} \w_i^{\ b' a''} 
(\d_{a' b'}-i\vare_{a' b'}) \g_{a''} \} \d\q .
\eea
As discussed in the previous section, at the level of the quadratic
contributions to the path-integrals all terms should be evaluated at the 
classical values of the backgound fields. Therefore, the factor $2T_S RV^{-1/3}$
can be absorbed by a redefinition of the $\d\q$'s. Next we use the relation
\be
\int \cD \d \q \, e^{\int d^2 \s \d \q^{\dagger} \cO \d \q} \propto \det \cO .
\label{fpathint}
\ee
Note, however, that when going to Euclidean space, we have doubled the number
of fermion degrees of freedom. Therefore, one must actually integrate over only
one half of 
these degrees of freedom. This amounts to taking the square-root of the
determinant on the right-hand side of (\ref{fpathint}). Hence, we conclude
that
\be
\int \cD \d \q \, e^{- \S_2^{\q}} \propto \sqrt{\det \Oslash_3 } ,
\ee
where
\bea
\Oslash_3  &=&  \sqrt{g} \g_{a''} g^{ij} \{ e_{m''}^{\ a''} \del_j y^{m''} 
[ \del_i -i \w_i^{\ a' b'} \vare_{a' b'} + \w_i^{\ b'' c''} \g_{b'' c''}] \nn \\
& & \ \ \ \ \ \ \ \ \ \ + \d_j^{m'}e_{m'}^{\ a'} \w_i^{\ b' a''} 
(\d_{a' b'}-i\vare_{a' b'}) \} . \label{O3}
\eea

Collecting the results we have obtained thus far, two-point function
(\ref{pathint5}) can now be written as
\bea
\langle \l^I (y^u_1) \l^J (y^u_2) \rangle \; &\propto &
\frac{\sqrt{\det \Oslash_3 }}{\sqrt{\det \cO_1} \sqrt{\det \cO_2}} \cdot
\int \cD \l \, e^{-\int d^4 y \sum_{K=1}^{h^{1,1}}
\l^K \delslash  \l^K} \l^I (y^u_1) \l^J (y^u_2) \nn \\ 
& & \cdot \int d^4 x\, e^{- \frac{T}{2}\sum_{I=1}^{h^{1,1}}\w_I T^I (x)}
\sum_{L,M=1}^{h^{1,1}} \w_L \w_M \, \l^L (x) \l^M (x) \nn \\ 
& &  \cdot \int \cD \w \,  e^{-S_{0W\!Z\!W}}. \label{pathint7}
\eea
It, therefore, remains to evaluate the $\int \cD \w \, e^{-S_{0W\!Z\!W}}$ 
factor in the path-integral, which we now turn to.

\section{The Wess-Zumino-Witten Determinant:}

In this section, we will discuss the $E_8 \times E_8$ Wess-Zumino-Witten
part of the action, its quadratic expansion and one loop determinant. 
Recall from (\ref{boswzw}) that the relevant action is 
\bea
S_{0W\!Z\!W} &=& - \frac{1}{8 \pi} \int_{\CC} 
 d^2 \s \; \mbox{tr}[\frac{1}{2} \sqrt{g}g^{ij} ( \bar{\w}_i - \bar{A}_i )
 \cdot ( \bar{\w}_j - \bar{A}_j ) + i \vare^{ij} \bar{\w}_i \bar{A}_j ]\nn \\
& & + \frac{1}{24 \pi} \int_{\B} d^3 \hat{\s} i \hat{\vare}^{\hi\hj\hat{k}}
\Omega_{\hat{k}\hj\hi} (\bar{\w}') , \label{8.1}
\eea
where $\bar{\w}=\bar{g}^{-1}d\bar{g}$ is an $E_8 \times E_8$ Lie algebra valued
one-form and $\bar{g}$ is given in (\ref{e8*e8}). 
In order to discuss the equation of motion and the chirality of this 
action, it is convenient to use the complex coordinates $z=\s^0 +i\s^1$,
$\bar{z}=\s^0 -i\s^1$ on $\CC$ and to 
define the complex components of $\bar{A}$ by 
$\bar{A}=\bar{A}_z dz + \bar{A}_{\bar{z}}d\bar{z}$.
Then action (\ref{8.1}) can be written as
\bea
S_{0W\!Z\!W}&=&-\frac{1}{8\pi}\int_{\CC} d^2 z \;\mbox{tr} \left( \bar{g}^{-1}
\del_z \bar{g} \bar{g}^{-1} \del_{\bar{z}} \bar{g} - 2 \bar{A}_z 
\bar{g}^{-1} \del_{\bar{z}} \bar{g} +
\bar{A}_{\bar{z}} \bar{A}_z \right) \nn \\
& & + \frac{1}{24 \pi} \int_{\B} d^3 \hat{\s} i \hat{\vare}^{\hi\hj\hat{k}}
\Omega_{\hat{k}\hj\hi} (\bar{\w}') . \label{WZWA}
\eea
It is useful to define the two $E_8 \times E_8$ currents
\be
J_z = (D_z \bar{g}) \bar{g}^{-1}, \ \ \ \ \ \ \ \ \ \ \ \ \ \ \ 
J_{\bar{z}} = \bar{g}^{-1} D_{\bar{z}} \bar{g} ,
\ee
where $D_z$ and $D_{\bar{z}}$ are the $E_8 \times E_8$ covariant derivatives.
It follows from (\ref{WZWA}) that the $\bar{g}$ equations of motion are
\be
\del_{\bar{z}} J_z = 0 , \label{geom1}
\ee
or, equivalently,
\be
D_z J_{\bar{z}} + F_{z \bar{z}} = 0 , \label{geom2}
\ee
where $F_{z \bar{z}}$ is the $E_8 \times E_8$ field strength which, generically,
is non-vanishing.

In order to perform the path-integral over $\w$, it is necessary to fix any
residual gauge freedom in the $\w$ fields. Recall from the discussion in 
Section 3 that the entire action is invariant under both local gauge and
modified $\k$-transformations $\d_{\BbL}$ and $\D_{\hat{k}}$, respectively.
It follows from (\ref{gtransf}) and (\ref{gkappa}) that 
\be
\d_{\hat{\k}} \bar{g} = \bar{g} i_{\hat{k}} \bar{\BbA} .
\ee
It is not difficult to show that using this transformation, one can choose a
gauge where
\be
J_z = 0. \label{Jziszero}
\ee
Henceforth, we work in this chiral gauge. Note that this is consistent with the
equations of motion (\ref{geom1}) and (\ref{geom2}). Indeed, (\ref{geom1}) is
now vacuous, being replaced by (\ref{Jziszero}) itself.

Thus, the on-shell theory we obtain from the gauged Wess-Zumino-Witten action 
is an $E_8 \times E_8$ chiral current algebra 
at level one. The level can be read off from the coefficient of the Chern-Simons
term in (\ref{WZWA}). We would now like to evaluate the Wess-Zumino-Witten
contribution to the path-integral using a saddle-point approximation. To do 
this, we should expand $\bar{g}$ as small fluctuations 
\be
\bar{g} = \bar{g}_0 + \d \bar{g}
\ee
around a classical solution $\bar{g}_0$ of (\ref{Jziszero}). However, it is 
clearly rather difficult to carry out the quadratic expansion and evaluate the 
determinant in this formalism. Luckily, there is an equivalent theory which is 
more tractable in this regard, which we now describe.

To start the discussion, set all $E_8 \times E_8$ background gauge fields to
zero. It is well known that this is equivalent, on
an arbitrary Riemann surface, to an $SO(16)\times SO(16)$ theory of free
fermions. The partition function of the $E_8 \times E_8$ theory can be 
obtained by computing the partition function of the free fermion theory,
where one sums over all spin structures of each $SO(16)$ factor on the
Riemann surface \cite{GSW1}. We now want to turn on an arbitrary non-vanishing $E_8\times
E_8$ gauge field background. Here one meets some difficulty. Since only the 
$SO(16)\times SO(16)$ ($\subset E_8 \times E_8$) symmetry is manifest in the
free fermion theory, only background gauge fields associated with this subgroup
can be coupled to the fermions in the usual way. The coupling of $E_8\times
E_8$ gauge backgrounds with a structure group larger than $SO(16) \times
SO(16)$ is far from manifest and cannot be given a Lagrangian description.
Happily, for our purposes, it is sufficient to restrict the gauge field
backgrounds to lie within the $SO(16)\times SO(16)$ subgroup of
$E_8 \times E_8$. The reason is that, as explained in a number of papers
\cite{B27,B15,B5,B4,B3}, realistic heterotic M-theory requires gauge bundles with structure
group of rank 4 or smaller within the observable $E_8$ factor. In addition,
the structure group in the hidden $E_8$ factor can always be chosen to be of
rank 4 or less. Hence, realistic heterotic M-theory models can generically be
chosen to have gauge field backgrounds within the $SO(16)\times SO(16)$
subgroup of $E_8 \times E_8$. Henceforth, we consider only such restricted
gauge field backgrounds. With this in mind, we can now write the action for
$SO(16)\times SO(16)$ fermions coupled to background gauge fields. It is
given by \cite{VDP,RS1,RS2,RST,ST}
\be
S_{\psi} = \int_{\CC} d^2 \s ( \bar{\psi}^a \Dslash_A^{ab} \psi^b
+ \bar{\psi}^{a'} \Dslash_{A'}^{a'b'} \psi^{b'} ),
\ee
where $\psi^a$, $\psi^{a'}$ denote the two sets of $SO(16)$ fermions with
$a,a'=1,\ldots,16$ and
\be
\Dslash_A^{ab} = \sqrt{g} \t^i (D_i \d^{ab} - A_i^{ab}), \ \ \ \ \ \ \ \ \ 
\ \ \ \ \ \ \Dslash_{A'}^{a'b'} = \sqrt{g} \t^i (D_i \d^{a'b'} - A_i^{a'b'})
\label{DA}
\ee
are the covariant derivatives on $\CC$ with $A_i^{ab}$, $A_i^{a'b'}$ the two
sets of $SO(16)$ background gauge fields. Recall that $\t^i$ are the Dirac
matrices in two-dimensions. It follows from the above discussion
that we can write
\be
\int \cD \w \, e^{-S_{0W\!Z\!W}} \, \propto \int \cD \psi^a \cD \psi^{a'} \,
e^{-S_{\psi}} , \label{wzwpathint}
\ee
where the gauge fixing of variable $\w$ described by (\ref{Jziszero}) is 
inherent in the $\psi^a$, $\psi^{a'}$ formalism, as we will discuss below. 
The equations of motion are given by
\be
\Dslash_A^{ab} \psi^b = 0 , \ \ \ \ \ \ \ \ \ \ \ \ \ \ \ \ 
\Dslash_{A'}^{a'b'} \psi^{b'} = 0 . \label{eomforpsi}
\ee
We now expand
\be
\psi^a = \psi_0^a + \d \psi^a , \ \ \ \ \ \ \ \ \ \ \ \ \ \ \ 
\psi^{a'} = \psi_0^{a'} + \d \psi^{a'}
\ee
around a solution $\psi_0^a$, $\psi_0^{a'}$ of (\ref{eomforpsi}) and consider
terms in $S_{\psi}$ up to quadratic order in the fluctuations 
$\d\psi^a$, $\d\psi^{a'}$. We find that
\be
S_{\psi} = S_{0\psi} + S_{2\psi},
\ee
where
\be
S_{0\psi} = \int_{\CC} d^2 \s ( \bar{\psi}_0^a \Dslash_A^{ab} \psi_0^b
+ \bar{\psi}_0^{a'} \Dslash_{A'}^{a'b'} \psi_0^{b'} )
\ee
and
\be
S_{2\psi} = \int_{\CC} d^2 \s ( \d\bar{\psi}^a \Dslash_A^{ab} \d\psi^b
+ \d\bar{\psi}^{a'} \Dslash_{A'}^{a'b'} \d\psi^{b'} ) . \label{S2psi}
\ee
The terms linear in $\d\psi$ vanish by the equations of motion. It follows 
immediately from (\ref{eomforpsi}) that $S_{0\psi} = 0$. Then, using
(\ref{fpathint}), one finds from (\ref{S2psi}) that
\be
\int \cD \psi^a \cD \psi^{a'} \, e^{-S_{\psi}} \, \propto \sqrt{\det \Dslash_A} 
\sqrt{\det \Dslash_{A'}}. \label{gaugpathint}
\ee
Note, again, that by going to Euclidean space we have doubled the number of 
fermionic degrees of freedom. Therefore, one must actually integrate over only
one half of these degrees of freedom. This requires the square-root of the
determinants to appear in (\ref{gaugpathint}).
It is important to discuss how the chiral gauge fixing condition 
(\ref{Jziszero}) is manifested in the $\psi^a$, 
$\psi^{a'}$ formalism. Condition (\ref{Jziszero})
imposes the constraint that $\bar{g}$ couples only to the $\bar{A}_z$ component 
of the gauge fields and not to $\bar{A}_{\bar{z}}$. It follows that in 
evaluating $\det \Dslash_A \det \Dslash_{A'}$, we should keep only the 
$\bar{A}_z$ components of the gauge
fields. That is, we should consider the Dirac determinants of
$SO(16)\times SO(16)$ holomorphic vector bundles on Riemann surface $\CC$.
Gauge fixing condition (\ref{Jziszero}) also imposes a constraint on the 
definition of determinants $\det \Dslash_A \det \Dslash_{A'}$ as follows. 
Consider one of the $SO(16)$ Dirac operators, say $\Dslash_A$. Recall that on 
the Euclidean space
$\CC$, each spinor $\psi$ is a complex two-component Weyl spinor
\be
\psi = \left( \ba{c} \psi_+ \\ \psi_- \ea \right) .
\ee
Rescaling this basis to
\be
\left( \ba{c} \psi_+ \\ \psi_- \ea \right) = 
\left( \ba{c} (g_{z\bar{z}})^{-1/4} \tilde{\psi}_+ \\ 
(g_{z\bar{z}})^{1/4} \tilde{\psi}_- \ea \right)
\ee
and using the standard representation for $\t^0,\t^1$ then, locally, one can
write
\be
\Dslash_A = \left( \ba{cc} 0 & D_{-A} \\ D_{+A} & 0 \ea \right) , 
\label{DAmatrix}
\ee
where
\be
D_{-A} = (g_{z\bar{z}})^{3/4} \left( (g_{z\bar{z}})^{-1/2} \frac{\del}{\del z} 
(g_{z\bar{z}})^{1/2} - A_z \right) , \ \ \ \ \ \ \ \ \ \ \ \ \ \ \ 
D_{+A} = (g_{z\bar{z}})^{1/4} \frac{\del}{\del \bar{z}} . \label{D-AD+A}
\ee
Since the operator $\Dslash_A$ must be Hermitean, it follows that 
$D_{+A}=D_{-A}^{\dagger}$. Now, in addition to disallowing any coupling to
$A_{\bar{z}}$, gauge condition (\ref{Jziszero}) imposes the constraint that
\be
\psi_+^a = 0 \label{psi+iszero}
\ee
for all $a=1,\ldots,16$. Then, using the fact that
\be
\det \Dslash_A = \sqrt{\det (\Dslash_A)^2}
\ee
and gauge condition (\ref{psi+iszero}), we see that the proper definition of
the determinant is
\be
\det \Dslash_A = \sqrt{\det D_{-A}^{\dagger} D_{-A}} . \label{detDA}
\ee
In this paper, it is not necessary to determine the exact value of 
$\det \Dslash_A$.
We need only compute whether it vanishes or is non-zero, and the conditions 
under which these two possibilities occur. To do this, we must examine the
global properties of the holomorphic vector bundle. As we did in Section 7,
we will, henceforth, restrict
\be
\CC = \BbC \BbP^1 = S^2 .
\ee
With this restriction, the condition for the vanishing of $\det \Dslash_A$ can
be given explicitly, as we now show.

It follows from (\ref{psi+iszero}) that the chiral fermions realizing the 
$SO(16)$ current algebra are elements of the negative chiral spinor line 
bundle $S_-$ of the sphere. Note from (\ref{DAmatrix}) that $D_{-A}$
is the part of the Dirac operator which acts on $S_-$. With respect to a
non-trivial $SO(16)$ gauge bundle background $A$, the complete operator we
should consider is 
\be
D_{-A}: S_- \otimes A \to S_+ \otimes A  , \label{D-map}  
\ee
where $S_+$ denotes the positive chiral spinor bundle on the sphere. This is
the global description of the local $D_{-A}$ operator defined in 
(\ref{DAmatrix}) and (\ref{D-AD+A}).\footnote{To be even more precise,
globally $D_{-A}$ is a holomorphic section of the determinant line bundle
$\det (S_- \otimes A) \otimes \det (S_+ \otimes A)^{\star}$ over the gauge
fixed configuration space. Furthermore, definition (\ref{detDA}) corresponds
to the Quillen norm of this holomorphic section. For a careful definition of
the determinant of infinite dimensional spaces $S_- \otimes A$ and
$S_+ \otimes A$ and the Quillen norm, see \cite{AGetal,Freed}.} 
In order to have nonzero determinant $\det \Dslash_A$, it is necessary and 
sufficient
that $D_{-A}$ should not have any zero-modes. This follows from the fact that, 
for $SO(16) \times SO(16)$, the index theorem guarantees that
\be
\mbox{coker} D_{-A}^{\dagger} = \mbox{ker} D_{-A} .
\ee
In Appendix B, we discuss how to 
calculate the number of zero-modes of the $D_{-A}$ operator globally defined
in (\ref{D-map}). Since the number of zero-modes is invariant under
a smooth deformation of the bundle, we can choose a convenient bundle
for the purpose of the calculation. It is known that by a judicious
choice of the connection, every $E_8$ bundle can be reduced to 
a $U(1)^8$ bundle. Since the maximal torus of $E_8$ coincides with that of
$SO(16)$, every $U(1)^8$ bundle should have the form
\be
\oplus_{i=1}^{8}{\cal O}(m_i)\oplus {\cal O}(-m_i) ,  \label{sumofOs}
\ee
where ${\cal O}(m_i)$ stands for the $U(1)$ bundle on the sphere with
degree $m_i$. Let us now suppose that the bundle $A$ is written in the form of
(\ref{sumofOs}). It follows that operator $D_{-A}$ decomposes as
\be
D_{-A} = \oplus_{i=1}^{8}D_{-{\cal O}(m_i)}\oplus D_{- {\cal O}(-m_i)} .
\label{D-Adecomp} 
\ee
In Appendix B, we show that the number of zero-modes of the factor
\be
D_{- {\cal O}(m)}: S_- \otimes {\cal O}(m) \to S_+ \otimes {\cal O}(m) 
\ee
is $m$ if $m$ is nonegative and zero otherwise. From this fact, we can see that
if any of the $m_i$ in (\ref{sumofOs}) is nonzero, then either 
$D_{- {\cal O} (m_i)}$ or $D_{- {\cal O} (-m_i)}$ has zero-modes 
since either $m_i$ or $-m_i$ must be positive. We conclude from this and
expression (\ref{D-Adecomp}) that $D_{-A}$ will have at least one zero-mode,
and hence $\det \Dslash_A$ will vanish, if and only if at least one $U(1)$ 
sub-bundle ${\cal O}(m_i)$ has degree $m_i \neq 0$. That is, $\det \Dslash_A 
= 0$
if and only if holomorphic bundle $A$, restricted to the sphere $\CC = S^2$
and denoted by $A \mid_{\CC}$, is non-trivial. Clearly, the exact same 
conclusions apply to the other $SO(16)$ determinant $\det \Dslash_{A'}$ as well.

It is obviously very important to know, within the context of realistic
heterotic M-theory vacua, when the restriction of the $SO(16)\times SO(16)$
gauge bundle $A \oplus A'$ to $\CC = S^2$, denoted $A \mid_{\CC} \oplus
A' \mid_{\CC}$, is non-trivial, in which case this
curve produces no instanton contributions to the superpotential $W$, and
when $A \mid_{\CC} \oplus A' \mid_{\CC}$ is trivial, in which case it gives
a non-zero contribution to $W$. We have shown that both situations are
possible, the conclusion depending on the explicit choice of the gauge bundle,
the specific Calabi-Yau threefold and the choice of the isolated sphere 
within a given Calabi-Yau threefold. These results will 
be presented elsewhere \cite{LOPR}.

\section{Final Expression for the Superpotential:}

We are now, finally, in a position to extract the final form of the 
non-perturbative superpotential from the fermion two-point function. 
Combining the results of the previous section with expression (\ref{pathint7}),
we find that
\bea
\langle \l^I (y^u_1) \l^J (y^u_2) \rangle \; &\propto &
\frac{\sqrt{\det \Oslash_3 }}{\sqrt{\det \cO_1} \sqrt{\det \cO_2}} \cdot
\sqrt{\det \Dslash_A} \sqrt{\det \Dslash_{A'}} \nn \\
& & \cdot \int \cD \l \, e^{-\int d^4 y \sum_{K=1}^{h^{1,1}}
\l^K \delslash  \l^K} \l^I (y^u_1) \l^J (y^u_2)
\nn \\ 
& & \cdot \int d^4 x\, e^{- \frac{T}{2}\sum_{I=1}^{h^{1,1}}\w_I T^I (x)}
\sum_{L,M=1}^{h^{1,1}} \w_L \w_M \, \l^L (x) \l^M (x) .
\eea
Comparing this with the purely holomorphic part of the quadratic fermion term
in the four-dimensional effective Lagrangian (\ref{4Daction})
\be
(\del_I \del_J W) \l^I \l^J ,
\ee
we obtain
\be
W \propto \frac{\sqrt{\det \Oslash_3 }}{\sqrt{\det \cO_1} \sqrt{\det \cO_2}} 
\cdot \sqrt{\det \Dslash_A} \sqrt{\det \Dslash_{A'}} \, \cdot e^{-\frac{T}{2}
\sum_{I=1}^{h^{1,1}}\w_I T^I}.
\label{final}
\ee
In this expression, the dimensionless fields $T^I$ correspond to the 
$(1,1)$-moduli and the volume modulus of $S^1/\Z_2$. The $\w_I$ are 
dimensionless coefficients defined by
\be
\w_I = \frac{1}{v_{\CC}} \int_{\CC} d^2 z \, \w_{Iz\bar{z}} ,
\ee
where $v_{\CC}$ is the volume of curve $\CC$ and $\w_{Iz\bar{z}}$ is the
pullback (\ref{wIzz}) to the holomorphic curve $\CC$ of the $I$-th harmonic
$(1,1)$-form on $CY_3$. $T$ is a dimensionless parameter given by
\be
T = T_M \pi \rho \ v_{\CC},
\ee
with $T_M$ the membrane tension and $\pi \rho$ the $S^1/\Z_2$ interval length.
The operators $\cO_1,\cO_2$ and $\Oslash_3 $ are presented in (\ref{Os}) and 
(\ref{O3}), respectively. The operator $\Dslash_A$ and its determinant 
$\det \Dslash_A$
are defined in (\ref{DA}), (\ref{DAmatrix}), (\ref{D-AD+A}) and (\ref{detDA}).
This determinant and, hence, the superpotential $W$ will be non-vanishing if
and only if the pullback of the associated $SO(16)$ bundle $A$ to the curve
$\CC$ is trivial. The same results apply to $\Dslash_{A'}$, $\det \Dslash_{A'}$
and the
pullback of the other $SO(16)$ bundle $A'$ to $\CC$. All the determinants
contributing to $W$ are non-negative real numbers. We emphasize that $W$ given
in (\ref{final}) is the contribution of open supermembranes wrapped once
around $\CC \times S^1/\Z_2$, where $\CC = S^2$ is a sphere isolated in the
Calabi-Yau threefold $CY_3$. The existence of supermembranes multiply wrapped
around $\CC$, and the computation of their superpotential, is not 
straightforward. However, we expect that the appropriate contribution to
the superpotential of a supermembrane wrapped once around $S^1/\Z_2$ and $n$
times around $\CC$ is
\be
e^{- \frac{nT}{2} \sum_{I=1}^{h^{1,1}} \w_I T^I} .
\ee
Further generalizations and discussions of the complete open supermembrane
contributions to the non-perturbative superpotential in heterotic M-theory
will be presented elsewhere \cite{LOPR}.

\appendix

\section{Notation and Conventions:}

We use a notation such that symbols and indices without hats represent 
fields in the ten-dimensional fixed hyperplanes of \HW theory 
(as well as the two-dimensional heterotic string theory), 
while hatted indices relate to quantities of eleven-dimensional bulk space
(and the three-dimensional open membrane theory). 

\subsection*{Bosons:}
    
For example,
\be
X^M, \ \ \ M=0,1,\ldots,9, \ \ \ \ \mbox{and} \ \ \ \ \hat{X}^{\hat{M}},
\ \ \ \hat{M}=\hat{0},\hat{1},\ldots,\hat{9},\hat{11}, \label{app1}
\ee
are, respectively, the coordinates of ten- and eleven-dimensional spacetimes. 
We do not change notation when switching from Minkowskian signature to 
Euclidean signature.

Eleven-dimensional space is, by assumption, given by
\be
M_{11} = R_4 \times CY_3 \times S^1/\Z_2 ,
\ee
while the ten-dimensional space obtained by compactifying it on $S^1/\Z_2$ is, 
clearly,
\be
M_{10} = R_4 \times CY_3 .
\ee
The membrane world volume $\Si$ is decomposed as
\be
\Si = \CC \times S^1/\Z_2 ,
\ee
where the (two-dimensional) curve $\CC$ lies within $CY_3$.

The two dimensional heterotic string theory is represented by fields of the
worldsheet coordinates $\s^i$, with $i=0,1$. Bosonic indices of ten-dimensional
spacetime are split into indices parallel to the worldsheet
($m' = 0,1$) and indices perpendicular to it ($m''=2,\ldots,9$).
The space normal to the worldsheet is an eight-dimensional space. Since
it is assumed that the worldsheet $\CC$ is contained in the Calabi-Yau 
three-fold $CY_3$, these eight directions $y^{m''}$ can be split in two sets 
of four, the first being directions of $CY_3$ but perpendicular to $\CC$, 
denoted by
$y^U$, with $U=2,3,4,5$, and the second being directions normal to $CY_3$,
that is, directions of $R_4$, denoted by $y^u$, with $u=6,7,8,9$. 

Coordinates of $CY_3$ are denoted by
\be
\breve{y}^{\breve{U}} = (X^{m'},y^U) , \ \ \ \ \ \ \ \mbox{with} \ \ \ 
\breve{U}=0,1,2,3,4,5,\ \ \ \ m'=0,1,
\ee
or, using the complex structure notation, 
\be
\breve{y}^m, \ \ \ \breve{y}^{\bar{m}}, \ \ \ \ \ \ \ \ \ \ \ \
m= 1,2,3, \ \ \ \ \bar{m}=\bar{1},\bar{2},\bar{3}.    \label{app6}
\ee

The bosonic indices in (\ref{app1})-(\ref{app6}) are coordinate 
(or ``curved'') indices. The corresponding
tangent space (or ``flat'') indices are given in the following table,

\begin{center}
\begin{tabular}{||c|c|c|c|c|c||} \hline
$M_{10}$ & $M_{11}$ & $\CC$ & $M_{\perp}$ & $R^4$ & $CY_{\perp}$ \\ \hline 
\hline
$M,N$ & $\hat{M},\hat{N}$ & $m',n'$ & $m'',n''$ & $u,v$ & $U,V$ \\ \hline
$A,B$ & $\hat{A},\hat{B}$ & $a',b'$ & $a'',b''$ & $k,l$ & $K,L$ \\ \hline
\end{tabular}
\end{center}

\noindent where $M_{\perp}$ and $CY_{\perp}$ are subspaces of $M_{10}$ and 
$CY_3$ perpendicular to $\CC$, respectively.

\subsection*{Spinors:}
    
In ten-dimensional spacetime with Euclidean signature, the $32\times 32$ 
Dirac matrices $\G_A$ satisfy
\be
\{ \G_A, \G_B \} = 2 \eta_{A B}
\ee
or, with curved indices, (since $\G_A = e_A^{\ M} \G_M$)
\be
\{ \G_M, \G_N \} = 2 g_{M N}.
\ee
One defines ten-dimensional chirality projection operators 
$\frac{1}{2} (1\pm \tilde{\G})$,
where
\be
\G_{11} = - i \G_0 \G_1 \cdots \G_9.
\ee
A useful representation for $\G_A$ is given by the two-eight split
\be
\G_{A} = ( \t_{a'} \otimes \tilde{\g} , 1 \otimes \g_{a''} ) ,
\ee
where the two-dimensional Dirac matrices $\t_0,\t_1$ and their product defined
by $\tilde{\t}= - i\t_0 \t_1$ are explicitly given by
\be
\t_0 = \left( \ba{cc} 0 & 1 \\ 1 & 0 \ea \right) , \ \ \ 
\t_1 = \left( \ba{cc} 0 & -i \\ i & 0 \ea \right) , \ \ \ 
\tilde{\t} = \left( \ba{cc} 1 & 0 \\ 0 & -1 \ea \right). 
\ee
These ten-dimensional Dirac matrices are more explicitly written as
\be
\ba{cc}
\G_0 = \left( \ba{cc} 0 & \tilde{\g} \\ \tilde{\g} & 0 \ea \right) &
\G_1 = \left( \ba{cc} 0 & -i \tilde{\g} \\ i \tilde{\g} & 0 \ea \right) \\
\G_{a''} = \left( \ba{cc} \g_{a''} & 0 \\ 0 & \g_{a''} \ea \right) &
\G_{11} = \left( \ba{cc} \tilde{\g} & 0 \\ 0 & -\tilde{\g} \ea \right) \ea
\ee
where $\g_{a''}$ are $16\times 16$ Dirac matrices, and the product
\be
\tilde{\g} = \g_2 \g_3 \cdots \g_9
\ee
is used in the definition of eight-dimensional chirality projection operators 
$\frac{1}{2} (1\pm \tilde{\g})$.

Note that $\G_{11}^2 = 1$, $ \tilde{\g}^2 = 1 $, and $ \tilde{\t}^2 = 1$. 
In eleven-dimensions, the $32 \times 32$ Dirac matrices are given by
\be
\hat{\G}_{\hat{A}} = \G_{\hat{A}}, \ \ (\hat{A}=\hat{0},\hat{1},\ldots,\hat{9}),
\ \ \ \ \ \mbox{and} \ \ \ \ \ \hat{\G}_{\hat{11}} = \G_{11}.
\ee

\section{Dirac Operator on the Sphere:}

In this Appendix, we discuss the spinors, the chiral Dirac operator 
on the sphere and the number of zero modes of the Dirac operator.  
One nice thing on the sphere and generally on the Riemann surfaces 
is that we can identify a chiral spinor with a holomorphic bundle and 
the associated Dirac operator with a suitable differential operator 
on the corresponding holomorphic bundle. We follow \cite{AGetal} closely in
the 
exposition.  

When we consider a sphere ${\cal C}$, locally we can always write the metric as
\begin{equation}
ds^2=e^{2\phi} ( (d\s^0)^2+ (d\s^1)^2)=2g_{z \bar{z}}dz d\bar{z}
\label{eqa1}
\end{equation}
with $z=\s^0+i\s^1$ and $\bar{z}=\s^0-i\s^1$. When we cover ${\cal C}$ with such
complex 
coordinate patches, the transition functions on the overlaps are
holomorphic 
and 
thus define a complex structure. Using this complex structure, we can
divide 
one forms
\begin{equation}
T^{*}{\cal C}=T^{*(1,0)}{\cal C} \oplus T^{*(0,1)}{\cal C}
\end{equation}
according to whether they are locally of the form $f(z, \bar{z})dz$ or 
$f(z, \bar{z})d\bar{z}$. 

In order to introduce spinors we choose a local frame $e^a$
\begin{equation}
ds^2=\delta_{ab}e^a\otimes e^b.
\end{equation}
If ${\cal U}_{\alpha}, {\cal U}_{\beta}$ are two overlapping coordinate 
patches, the corresponding frames $e^a_{\alpha}, e^a_{\beta}$ are related by 
a local $SO(2)$ rotation 
\begin{equation}
e^a_{\alpha}=R^a_{\, b}(\theta)e^b_{\beta} .
\end{equation}
Hence, the spinor bundles have transition functions $\tilde{R}(\theta/2)$
such 
that 
$\tilde{R}(\theta/2)^2=R(\theta)$. 

One of the pleasant features of the sphere or, more generally, of the
Riemann 
surfaces is 
that we can describe spinors in terms of half-order differential. 
This is most easily done by choosing frames 
\begin{eqnarray}
e^w&=&e^0+ie^1=e^{\phi}dz \ \ \ \ \ \ \ \ \  {\rm for} \ \ \ \ T^{*(1,0)},\nn \\
e^{\bar{w}}&=&e^0-ie^1=e^{\phi} d\bar{z} \ \ \ \ \ \ \ \ \ {\rm for} \ \ \ \  
T^{*(0,1)}, 
\end{eqnarray}
where $\phi$ is the factor in (\ref{eqa1}). 
Then across patches ${\cal U}_{\alpha}, {\cal U}_{\beta}$ 
with coordinates $z^a_{\alpha}, z^a_{\beta}$, we have
\begin{equation}
e^{2\phi_{\alpha}}|dz_{\alpha}|^2=e^{2\phi_{\beta}}|dz_{\beta}|^2 ,
\end{equation}
which implies that
\begin{equation}
2\phi_{\alpha}=2\phi_{\beta}+\ln|\frac{dz_{\beta}}{dz_{\alpha}}|^2.
\end{equation}
Thus we have 
\be
e^w_{(\alpha)}=e^{i\theta}e^w_{(\beta)} ,
\ee
where
\be
e^{i\theta}=\frac{dz_{\alpha}}{dz_{\beta}}|\frac{dz_{\beta}}{dz_{\alpha}}|.
\ee
The left- and right-spinors $\psi_{\pm} \in  S^{\pm}$ transform 
as 
\begin{equation}
\psi_{\pm \alpha}=e^{\pm \frac{i\theta}{2}} \psi_{\pm
\beta}
\end{equation}
on the sphere. When we refer the spinors $\psi_{\pm }$ to the
frame 
$(e^w)^{\frac{1}{2}}, (e^{\bar{w}})^{\frac{1}{2}}$, the transformation
functions
are one-by-one unitary matrices. 

It is sometimes more convenient to consider the bundle $S^{\pm}$ as 
{\it holomorphic} 
bundles. These will have transition functions 
$(\frac{dz_{\alpha}}{dz_{\beta}})^{\frac{1}{2}}$ for $S^+$ 
and $(\frac{dz_{\alpha}}{dz_{\beta}})^{-\frac{1}{2}}$ for $S^-$. 
In the holomorphic category more appropriate local sections are the 
holomorphic 
half order differential $(dz_{\alpha})^{\frac{1}{2}}$. The relation between
the 
standard and 
the holomorphic description of spinors is given by
\begin{equation}
\psi_+(e^w)^{\frac{1}{2}}= \tilde{\psi}_+ (dw_{\alpha})^{\frac{1}{2}}.
\end{equation}
The holomorphic line bundle defined by $S_+$ will be denoted by $L$. 
As suggested by the notation, this bundle can be interpreted as a
holomorphic 
square root of the bundle of $(1,0)$ form,
\begin{equation}
T^{*(1,0)}{\cal C}=K= L^2, 
\end{equation}
where $K$ denote the canonical bundle. 
Once we have introduced the spinor bundle $L$, we can define tensor powers 
$L^n$ corresponding to the differentials $\psi (dz_{\alpha})^{\frac{n}{2}}$. 
Note that by raising and lowering indices with the metric (\ref{eqa1}), any
tensor can be decomposed into such differentials.

In the local coordinate (\ref{eqa1}), the covariant derivative for fields in 
$L^n$ is 
\be
\nabla^n_z : L^n \longrightarrow L^{n+2} ,
\ee
such that
\be
\nabla^n_z \tilde{\psi} = (g_{z\bar{z}})^{\frac{n}{2}}\frac{\partial}{\partial
z} (g_{z\bar{z}})^{\frac{n}{2}} \tilde{\psi}.
\ee
We can introduce a scalar product in $L^n$
\begin{equation}
<\tilde{\phi}|\tilde{\psi}>=\int d^2 \sigma \sqrt{g}
(g^{z\bar{z}})^{\frac{n}{2}} \tilde{\phi}^*
\tilde{\psi} . 
\label{eqa2}
\end{equation}
The operator $\nabla^n_z$ is the unique holomorphic connection on
$L^n$ 
compatible with the inner product (\ref{eqa2}). With respect to 
(\ref{eqa2}), 
the adjoint is 
\be 
\nabla_{n+2}^z=(\nabla^n_z)^{\dagger} : L^{n+2} \longrightarrow L^{n} ,
\ee
with
\be
\nabla_{n+2}^z \tilde{\psi} = g^{z\bar{z}}\frac{\partial}{\partial \bar{z}}
\tilde{\psi}.
\ee
One can check that $\nabla^{-1}_z : L^{-1} \longrightarrow L^{1}$
coincides 
with the Dirac operator $D_{-}:S_{-}\longrightarrow S_{+}$. Together with 
$\nabla_{1}^z:L^1 \longrightarrow L^{-1}$, they form a Dirac operator on
the 
sphere. 

Let ${\cal O}(m)$ be a holomorphic bundle on the sphere with degree $m$. 
On the sphere $L\simeq {\cal O}(-1)$ since the holomorphic cotangent bundle is 
${\cal O}(-2)$. 
Under this notation, $S_{-}\simeq {\cal O}(1)$. Note that 
\begin{eqnarray}
S_{+}&\simeq & (\Lambda^{even} T^{*(0,1)}{\cal C})\otimes L, \nn  \\
S_{-}&\simeq & (\Lambda^{odd} T^{*(0,1)}{\cal C})\otimes L  
\end{eqnarray}
and, using the metric (\ref{eqa1}), $T^{*(0,1)}{\cal C}$ is identified with 
the holomorphic tangent bundle ${\cal O}(2)$.  

If we consider 
\begin{equation}
D_{- (m)}: S_{-}\otimes {\cal O}(-m) \rightarrow S_{+} \otimes {\cal O}(-m),
\end{equation}
where the connection on ${\cal O}(-m)$ is induced from the metric, 
then one can see that this coincides with the operator 
$\nabla^{m-1}_z : L^{m-1} \longrightarrow L^{m+1}$. 

According to the Riemann-Roch theorem $D_{- (m)}$ or $\nabla^{m-1}_z$ has
$m$ zero modes if $m$ is non-negative and zero otherwise \cite{AGetal,WJ}

\end{document}